\documentclass{ieeeaccess}
\usepackage{cite}
\usepackage{amsmath,amssymb,amsfonts}
\usepackage{algorithmic}
\usepackage{graphicx}
\usepackage{textcomp}
\usepackage{soul}
\usepackage{subcaption}  
\usepackage[square,sort,comma,numbers]{natbib}
\usepackage{diagbox}
\usepackage{stfloats}
\usepackage{multirow}
\usepackage{float}
\usepackage{parskip}

\newcommand{\nplancktot}{$242$}

\newcommand{\nplanckcc}{$137$}
\newcommand{\nhi}{$148$}
\newcommand{\ntot}{$506$}
\usepackage{caption}
\DeclareCaptionLabelSeparator{custom}{.}
\renewcommand{\thetable}{\Roman{table}}
\def\BibTeX{{\rm B\kern-.05em{\sc i\kern-.025em b}\kern-.08em
    T\kern-.1667em\lower.7ex\hbox{E}\kern-.125emX}}
\begin{document}
\history{ Received June, 1, 2022, accepted July, 5, 2022, date of publication xxxx 00, 0000, date of current version xxxx 00}
\doi{10.1109/ACCESS.2022.3189646}

\title{MaLeFiSenta: Machine Learning for FilamentS Identification and orientation in the ISM}
\author{\uppercase{Dana Alina}\authorrefmark{1,2},
\uppercase{Adai Shomanov \authorrefmark{1}, and Sarah Baimukhametova}.\authorrefmark{3}}
\address[1]{Nazarbayev University, Kabanbay batyr ave, 53, 010000 Nur-Sultan, Kazakhstan}
\address[2]{IRAP, Universite de Toulouse, CNRS, UPS, CNES, 31400 Toulouse, France}
\address[3]{University of Padua, Via 8 Febbraio, 2, 35122 Padova, Italy}
\tfootnote{This work was supported by the Science Committee of the Ministry of Education and Science of the Republic of Kazakhstan, in the frame of the project "Study of the polarised emission of the interstellar medium of the Magellanic Clouds using big data analysis and machine learning", Grant No. AP08855858. DA and SB acknowledge the Nazarbayev University Faculty Development Competitive Research Grant Programme No110119FD4503.}

\markboth
{Author \headeretal: Preparation of Papers for IEEE TRANSACTIONS and JOURNALS}
{Author \headeretal: Preparation of Papers for IEEE TRANSACTIONS and JOURNALS}

\corresp{Corresponding author: Dana Alina (e-mail: dana.alina@nu.edu.kz).}

\begin{abstract}
Filament identification became a pivotal step in tackling fundamental problems in various fields of Astronomy. Nevertheless, existing filament identification algorithms are critically user-dependent and require individual parametrization. This study aimed to adapt the neural networks approach to elaborate on the best model for filament identification that would not require fine-tuning for a given astronomical map. First, we created training samples based on the most commonly used maps of the  interstellar medium obtained by Planck and Herschel space telescopes and the atomic hydrogen all-sky survey HI4PI. We used the Rolling Hough Transform, a widely used algorithm for filament identification, to produce training outputs. In the next step, we trained different neural network models. We discovered that a combination of the Mask R-CNN and U-Net architecture is most appropriate for filament identification and determination of their orientation angles. We showed that neural network training might be performed efficiently on a relatively small training sample of only around 100 maps. Our approach eliminates the
parametrization bias and facilitates filament identification and angle determination on large data sets.
\end{abstract}


\titlepgskip=-15pt

\maketitle

\section{Introduction}
\label{sec:introduction}
Filaments are one of the main morphological structures of the baryonic compound of the Universe. They are present over many scales, both in the intergalactic and  interstellar medium.
The first evidence of the cosmic web structure in the distribution of galaxies was observed almost 50 years ago \citep{cosmo4,oort1983} and then reproduced in analytical models and simulations \citep[and references therein]{cosmo1,cosmo2,cosmo3,bond1996}.  
On the contrary, in the interstellar medium (ISM), filaments were first predicted by numerical simulations as elongated structures in density fields. 
Different authors ascribed their origin to compression and interpreted them as "cuts" through the sheets \citep{passot1987,stone1998,padoan1999a,padoan2001}. 
The presence of filaments in the ISM became irrefutable after the release of the \textit{Herschel} telescope images of interstellar dust emission \citep{andre2010,menshchikov2010}. Since then, filament studies constitute a buoyant topic because they coincide with sites of active star formation and act as evidence of the turbulent nature of the ISM and its interplay with the magnetic field and gravity \citep[to cite a few among many]{poidevin2011,palmeirim2013,li2013,planck2015-XXXII,federrath2016}.
In Solar Science, filament studies have also found its application  \citep{yuan2011automatic,xiao2016automatic} because filaments act as means of solar cycle detection \citep{barra2009fast}. 

Filament studies require their identification in a map, especially if one is interested in deriving statistical properties. Depending on the purpose, different authors adapted different pattern identification approaches or developed new techniques. 
Some methods are based on gradients, which are first-order derivatives \citep{soler2013}, while other methods use Hessian matrices \citep{schisano2014}, which involve second-order derivatives, to search for pixels with zero curvature to trace the filaments' crest. 
In the work by \cite{sousbie2013}, the author developed the so-called DisPerSE method, which combines the aforementioned approaches to detect voids, walls, and peaks in addition to filaments. This method was initially designed to be used on cosmological data but was further applied in the ISM studies \citep{doi2015}.
Another example is the \textit{Getsources} method developed by \cite{menshchikov2013} for the  \textit{Herschel} data. It constructs filtered decomposition of images over a range of spatial scales to be analyzed separately and reconstructs  the filaments.
There are also methods based on pattern recognition. The principle is to match a kernel of a given shape, usually a long rectangle, with an original image. Rolling Hough Transform (RHT, \cite{clark2014}), Template Matching (TM, \cite{juvela2016}) are such examples. 

The methods mentioned above can be virtually divided into two groups, depending on the purpose of the studies. The first group would encompass methods that enable tracing crests of the brightest structures (gradients, Hessian matrices, DisPerSE). The second group aims to identify the spatial extent of structures  (filaments having a width), regardless of their absolute intensity (RHT, TM).

Every method uses well-researched image processing techniques and has proven efficiency in specific studies. However, all  methods mentioned above are parameter-based and require parameter fine-tuning for every map.
Thus, detection of filaments in a large dataset by a given method is usually performed with a predefined set of parameters. The high dependence on a slight change in parametrization may cause bias in filament identification regarding their size or shape. 
Additionally, the existing methods commonly use advanced image processing procedures applied over the whole image field, which requires ample computational resources.
Moreover, a significant amount of time is spent on the parametrization of output maps by visual inspection. 
We propose a machine learning-based method for filament identification that solves the issue of manual parameter search and visual inspection. 

Neural networks are well suited to solve the filament identification problem. Recent developments in the field of deep learning, where object segmentation networks such as U-Net\cite{unet2015}, Mask R-CNN (Region-based Convolution Neural Network) \cite{he2017}, FastFCN (Fully Convolution Network)\cite{wu2019fastfcn}, Gated-SCNN (Gated-Shape Convolution Neural Network)\cite{takikawa2019gated}, DeepLab\cite{chen2017deeplab} provide an efficient and fast image segmentation results. Moreover, neural networks have already showed their efficiency in improving astronomical data and solve the problem of noise \citep{vojtekova2021,lauritsen2021}.

In astronomy, much effort is dedicated to methods of detection of specific morphological structures. For instance, in solar physics, the detection of bright points has been addressed with a combination of observation and simulation techniques\cite{crockett2010}. Analysis of granulation process on solar surface was treated using correlation tracking\cite{matloch2010,loptien2017}. 
Several studies have already used neural networks for filament identification. 
Authors in \cite{ahmadzadeh2019toward} proposed a neural network for solar filament segmentation, using a database of filaments detected by alternative methods. Their neural network is based on (R-CNN) model. 
Additionally, filament identification was researched in other domains, such as microscopy \cite{liu2018densely}. The authors proposed a densely connected stacked U-Net for filament segmentation in microscopy images. Similar work was performed by \cite{ozdemir2021automated} for automated and semi-automated enhancement, segmentation, and tracing of cytoskeletal networks in microscopic images. 

Our focus is to search for best neural network models that identify filamentary structures and their orientation angles in 2D maps, to apply it to the ISM studies. In particular, we are interested in detecting extended structures of a certain width and their orientation angles. For this purpose, we apply the RHT method to \textit{Planck}, \textit{Herschel} space telescopes, and the Effelsberg-Bonn and Parkes telescope survey data. This data contains a large number of interstellar dust filaments to produce training data sets. 
Nevertheless, it is worth noting that neural networks can be trained using filament identification methods other than the RHT and different training data sets.

This paper is organised as follows. First, we describe methods in Section \ref{sec:methods} and present the training datasets in Section \ref{sec:data}. We show and discuss our results in Sections \ref{sec:results} and \ref{sec:discussion} respectively. Finally, we summarise our work and provide quick tips for the network operation in Section \ref{sec:conclusion}.

\section{Method}
\label{sec:methods}
In this Section, we first outline the existing filament identification algorithm, the RHT, that was used to generate the training samples. Second, we describe the neural networks based on which we constructed our  models. Finally, we describe the principle of the goodness-of-fit mathematical measure that we use to assess the effectiveness of our results and during the preparation of the training samples.

\subsection{The Rolling Hough Transform}

\begin{figure}
    \centering
    \includegraphics[width = 0.7 \linewidth]{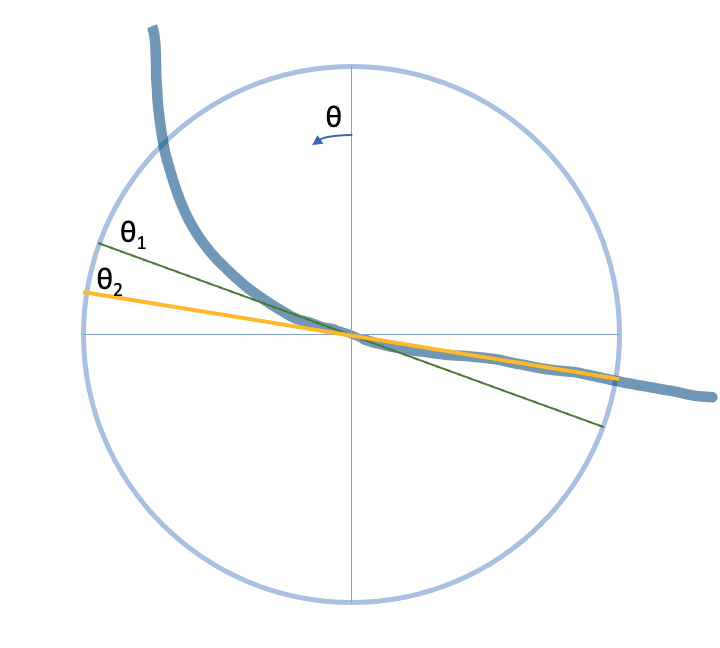}
    \caption{\quad Schematic representation of the RHT kernel scanning in a circular window centered at a pixel $(x,y)$. The blue curve represents he skeleton of the filament in the original image. The green and orange line represent thee kernel positions for different orientation angles $\theta_1$ and $\theta_2$ respectively. The width of the orange line is enhanced to show that for $\theta_2$ the histogram value is greater than the fixed threshold.}
    \label{fig:rht}
\end{figure}
The Rolling Hough Transform is based on the computer vision algorithm developed by Paul Hough in 1962 \citep{hough1962} to solve the problem of shape identification in 2D images. It employs parametrization from Cartesian coordinates $(x,y)$ to slope-intercept parameter space. Further, Duda and Hart \citep{duda1972} improved the method by replacing the slope-intercept space with angle-radius:
\begin{equation}
    \rho = x \cos{\theta} + y \sin{\theta} \, .
\end{equation}
This transformation ensures counting all pixels contributing to a pair $(\rho, \theta)$.
Duda and Hart's representation also facilitated the generalization of the Hough Transform from lines to various shapes such as ellipses, rectangles, or triangles. 

More recently, authors in \citep{clark2014} introduced the Rolling Hough Transform. The method consists of applying the Hough Transform at the same location inside a circular area of a fixed diameter, consequently in different directions. The number of pixels that are "on" the rotating rectangular kernel at each rotation position is stored. Thus, a histogram of orientation angles is built. The angle, for which a certain threshold is achieved for each pixel, is stored in the final output, as well as a measure of the cumulated intensity of the pixels. 

Practically, this is performed in the following way. First, a "top-hat" filter is applied on the image. Second, the resulting image is subtracted from the original image which provides skeletons of the structures. Third, a bitmap is created.  Finally, a rotating structuring element, called a kernel, is matched with the bitmap via the RHT. It is worth noting that for RHT, $\rho = 0$ because it is limited to a circular area, and the angle is retrieved by:
\begin{equation}
    \theta = \arctan(\frac{-x}{y}) \,.
\end{equation}
This last step is illustrated in Fig.~\ref{fig:rht}. For a more detailed representation of the RHT method, please refer to \cite{clark2014}.

Hence, RHT provides a mapping from the intensity of the 2D image to the orientation of detected structures. It is a practical method used in the studies of the ISM where relative orientation with respect to magnetic fields is often one of the main goals \citep{clark2014,malinen2016,alina2019}. However, as we have seen above, the RHT procedure requires repeating the Hough Transform at each pixel of the map making it computationally heavy.

We used the code available at GitHub\footnote{\underline{https://github.com/seclark/RHT}, based on \cite{clark2014}.} to generate our training samples described in Section \ref{sec:data}.

\subsection{Neural Networks architecture}
The input in our models is a 2D image, while the desired output consists of two maps: a mask representing filaments and a map with the values of the orientation angle of detected structures. We tested several neural network architectures such as the regular CNN model, the Mask-RCNN, the autoencoder model, the decision tree regression, and the U-Net model. We ran each of the four datasets through the models. We experimentally found that our task requires a two-step neural network that first identifies filaments and later determines the orientation angles. The best performing models are the Mask-RCNN and the U-Net model for filament identification and angle determination, respectively. The corresponding diagram is shown in Fig.~\ref{fig:arch}, while Section \ref{sec:results} describes the models in detail. 

\Figure[!t]()[width=\linewidth]{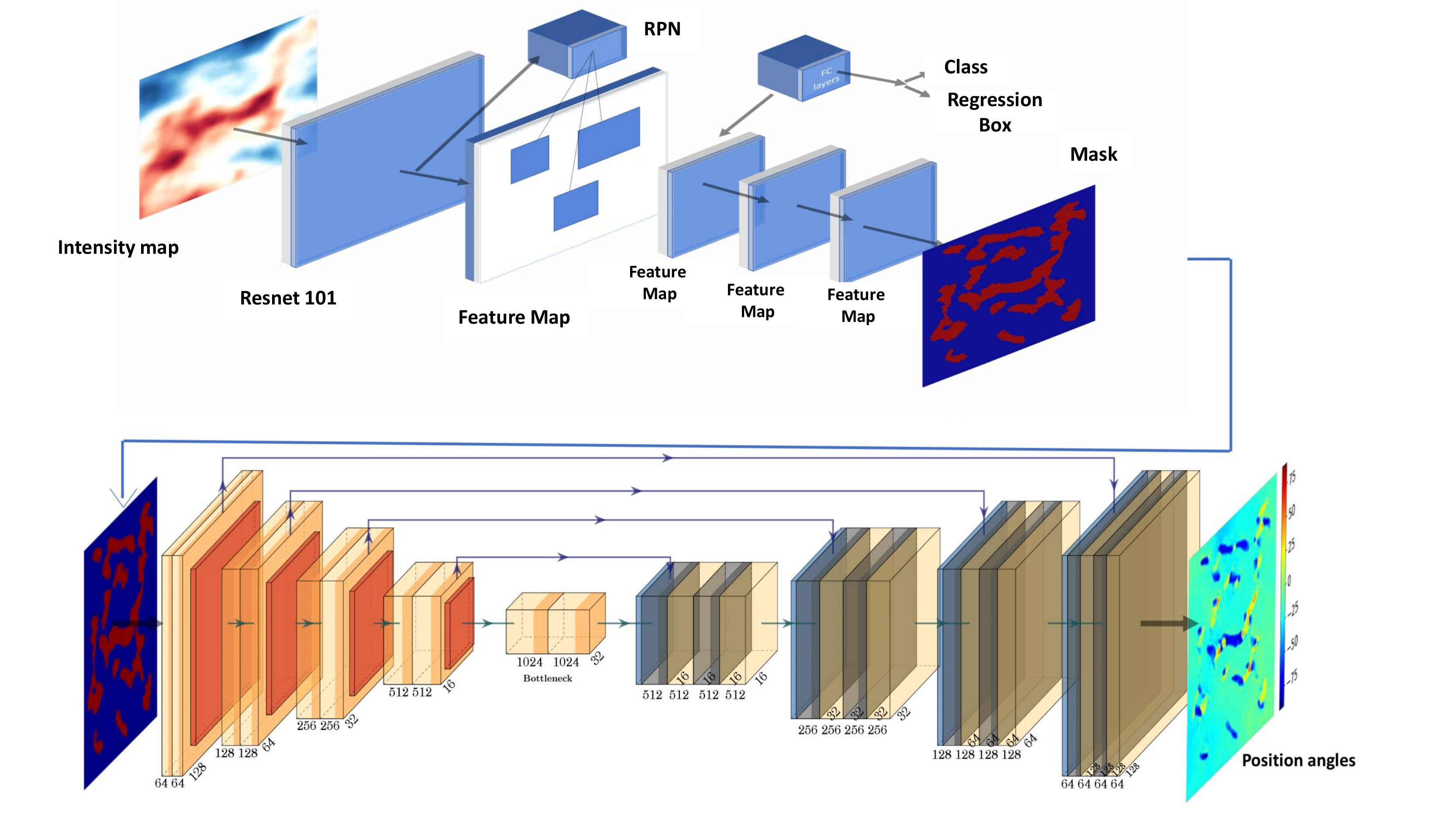}
   {\quad Diagram of the combined Mask-RCNN and U-Net model. Upper: Mask-RCNN, lower: U-Net.\label{fig:arch}}

\subsection{Measure of the structural similarity}
Minimization of human bias in filament identification is one of the principal aims of this work. 
The Mean Structural Similarity index (MSSIM) was introduced by \cite{wang2004} to measure similarity between images. 
\cite{green2017} proposed to use MSSIM in filaments analysis applied to the results of DisPerSE and FILFINDER filament identification algorithms which yield skeletons as output results. Here, we use MSSIM to quantify and compare the results of the neural networks approach to the results of the RHT. 
We used the publicly available\footnote{\underline{https://github.com/mubeta06/python/tree/master/signal\_processing/sp}} code written in Python. Below we shortly describe the main principles of the similarity index.

For a given pixel, the relationship between two images, or signals, $x$ and $y$, is characterized by "luminance", "contrast", and "structure" \citep{wang2004}. They are denoted as $\mathit{l}, \mathit{c}$, and $\mathit{s}$, respectively, and are, in fact, the mean intensity, the standard deviation, and the stored pattern, or the correlation between them:
\begin{eqnarray}
\mathit{l}(x,y) = \frac{2 \mu_x \mu_y + C_1}{ \mu^2_x + \mu^2_y + C_1} \, , \\
\mathit{c}(x,y) = \frac{2 \sigma_x \sigma_y + C_2}{\sigma^2_x + \sigma^2_y + C_2} \, , \\
\mathit{s}(x,y) = \frac{\sigma_{xy} + C_3}{\sigma_x \sigma_y + C_3} \, ,
\end{eqnarray}
where
\begin{eqnarray}
\mu_{x} = \frac{1}{N} \sum_{i=1}^{N} x_i \, , \hspace{0.5cm} \mu_y = \frac{1}{N} \sum_{i=1}^{N} y_i \, , \\
\sigma_{xy} = \Big( \frac{1}{N-1}\sum_{i=1}^{N}(x_i - \mu_x) \Big)(y_i - \mu_y) \Big)\, , \\
\sigma_x = \Big( \frac{1}{N-1}\sum_{i=1}^{N}(x_i - \mu_x)  \Big)^{1/2} \\
\sigma_y = \Big( \frac{1}{N-1}\sum_{i=1}^{N}(y_i - \mu_y)  \Big)^{1/2} 
\end{eqnarray}
are the mean values of signals $x$ and $y$ ($\mu_{x}$, $\mu_{y}$), their covariance ($\sigma_{xy}$), and the variances ($\sigma_x$, $\sigma_y$,  respectively). The constants $C_1,C_2$, and $C_3$ are much smaller than 1 and are introduced to avoid division by 0.
The local structural similarity index (SSIM) is given by the multiplication of the three above-cited parameters which results in the following expression:
\begin{equation}
\mathrm{SSIM} = \frac{ (2 \mu_x \mu_y + C_1) (2\sigma_{xy} + C_2)}{(\mu^2_x + \mu^2_y + C_1)(\sigma^2_x + \sigma^2_y + C_2)} \, ,
\end{equation}
given that $C_3 = C_2/2$ for simplification.

Then, the mean value of the SSIM index over an image provides a single value called MSSIM. It appears from \cite{brooks2008,green2017} that MSSIM primarily reflects the variation of the structure rather than luminance and contrast. For this reason, the metric is efficient for comparing the outputs of filament identification algorithms with the original images. 

MSSIM ranges from -1 to +1, where -1 means no similarity while +1 means perfect match. The higher the MSSIM value of the output image  containing identified filaments, the better the algorithm identifies critically essential structures of the image.

\section{Data}
\label{sec:data}
In our work, we collected approximately {\ntot} labelled images to train the neural networks.
We generated the datasets from the data from the following telescopes: \textit{Planck} and \textit{Herschel} space telescopes and Parkes and Effelsberg ground-based radio telescopes, the data from which is used the most in the analysis of ISM filaments.

\begin{figure}
    \centering
    %\begin{tabular}{cccc}
    \includegraphics[width = 0.22 \linewidth]{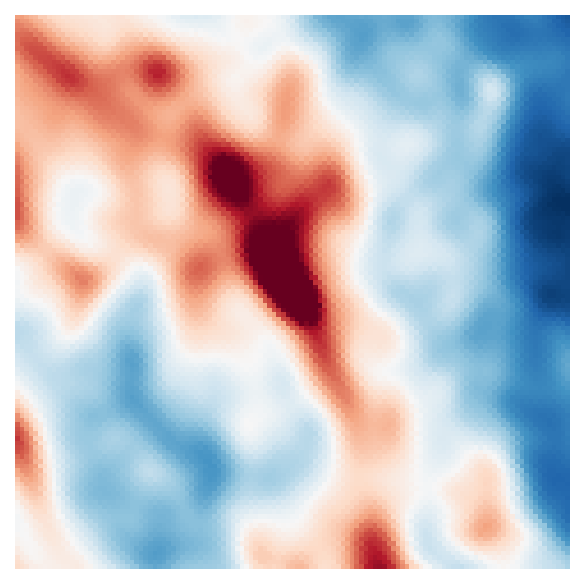} 
    \includegraphics[width = 0.22 \linewidth]{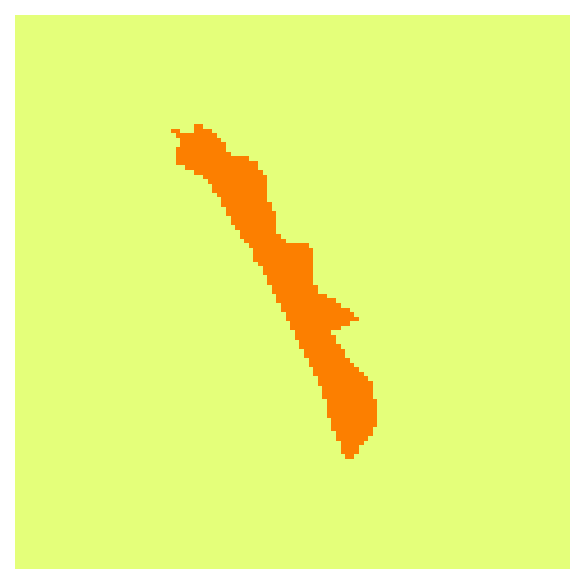} \hspace{0.3cm}
    \includegraphics[width = 0.22 \linewidth]{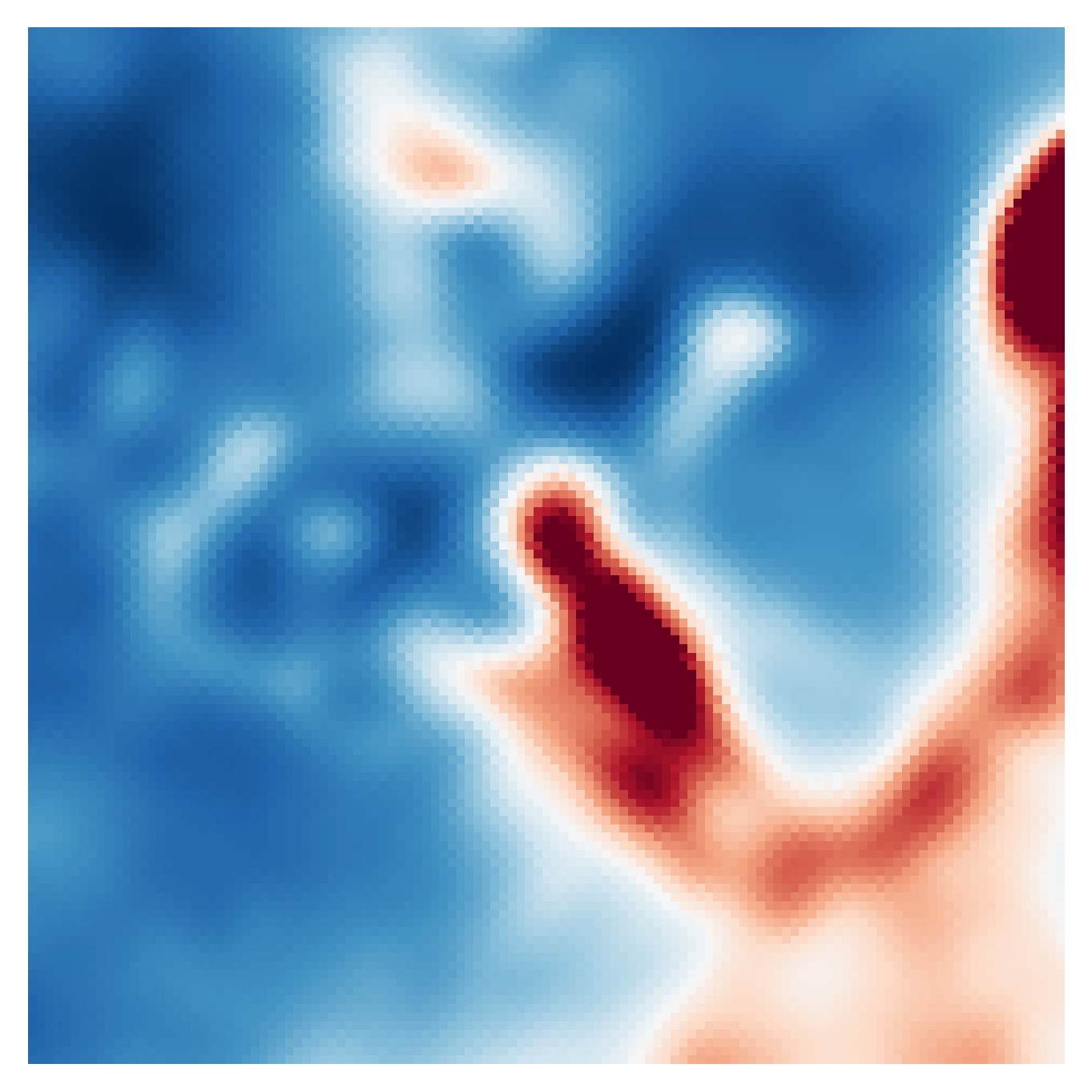} 
    \includegraphics[width = 0.22 \linewidth]{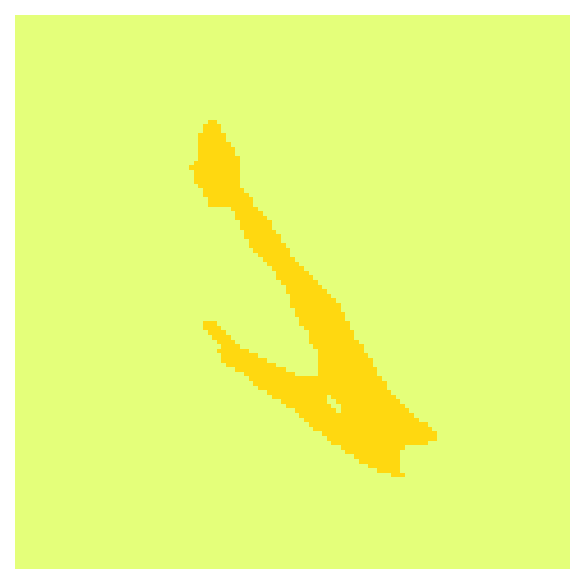} 
    \\
    \includegraphics[width = 0.22 \linewidth]{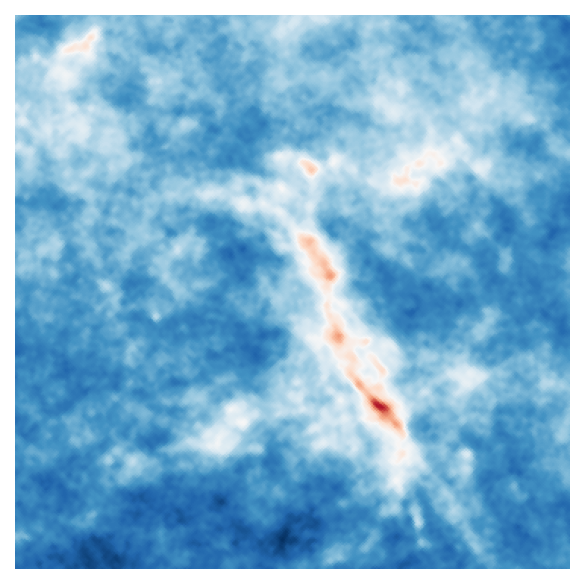} 
    \includegraphics[width = 0.22 \linewidth]{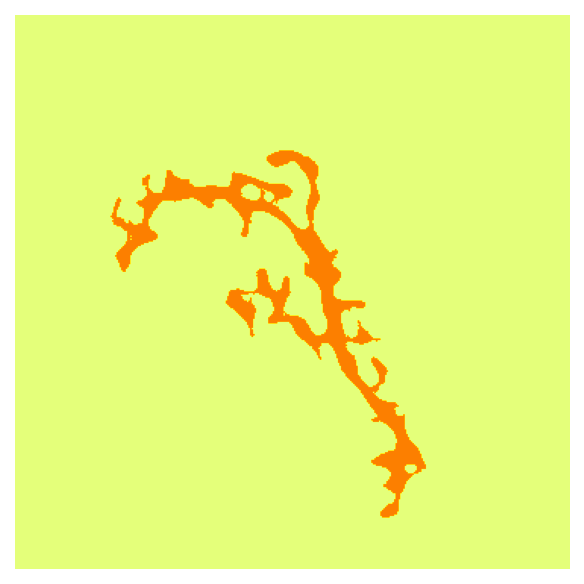} 
    \hspace{0.3cm}
    \includegraphics[width = 0.22 \linewidth]{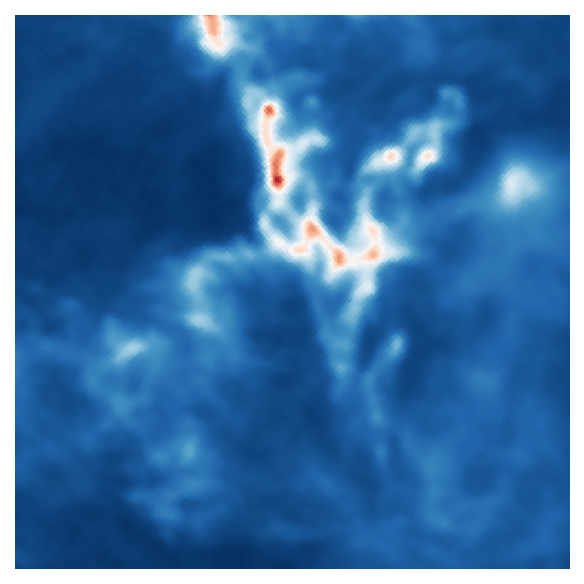} 
    \includegraphics[width = 0.22 \linewidth]{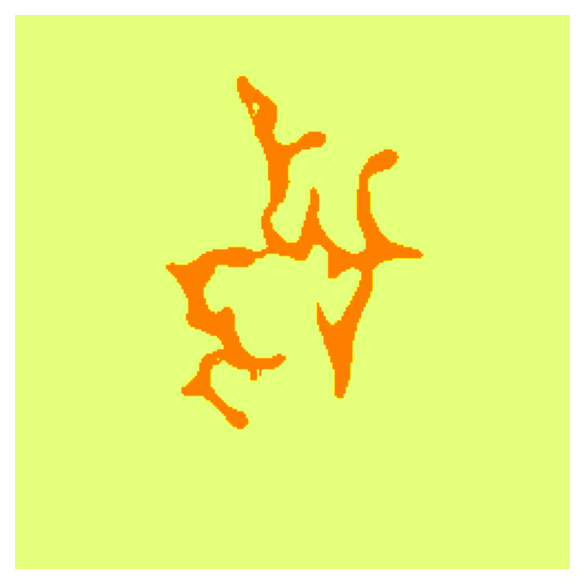} \\
    \includegraphics[width = 0.22 \linewidth]{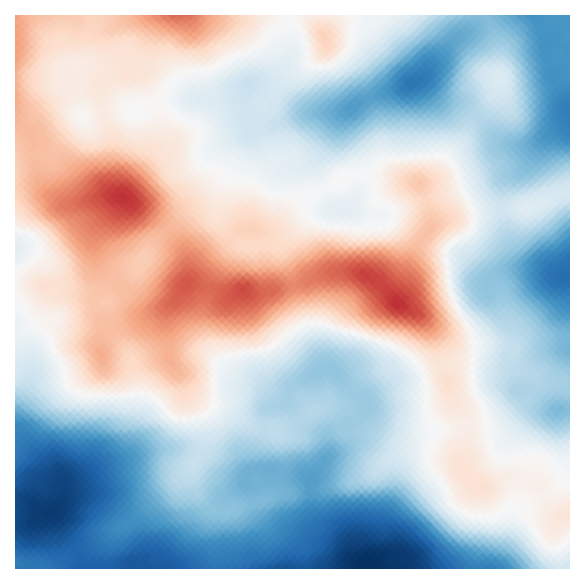} 
    \includegraphics[width = 0.22 \linewidth]{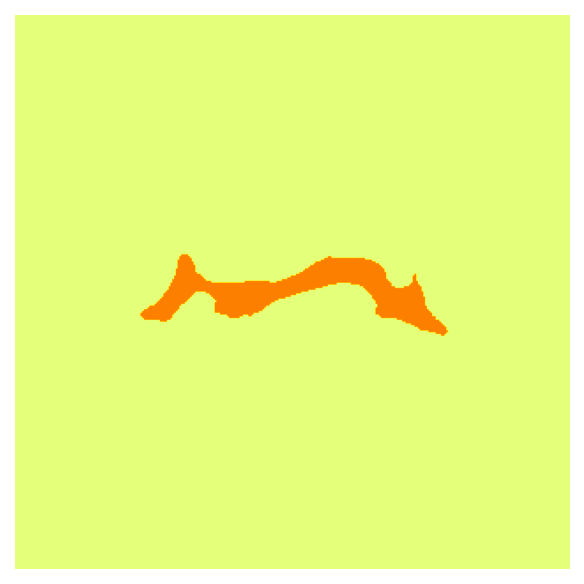} 
    \hspace{0.3cm}
    \includegraphics[width = 0.22 \linewidth]{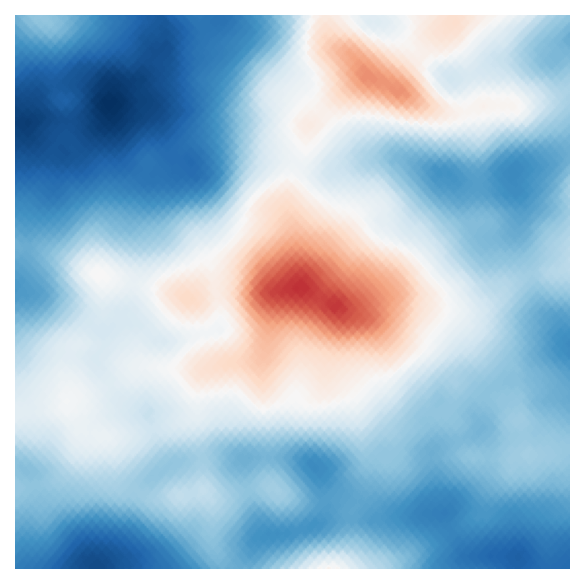} 
    \includegraphics[width = 0.22 \linewidth]{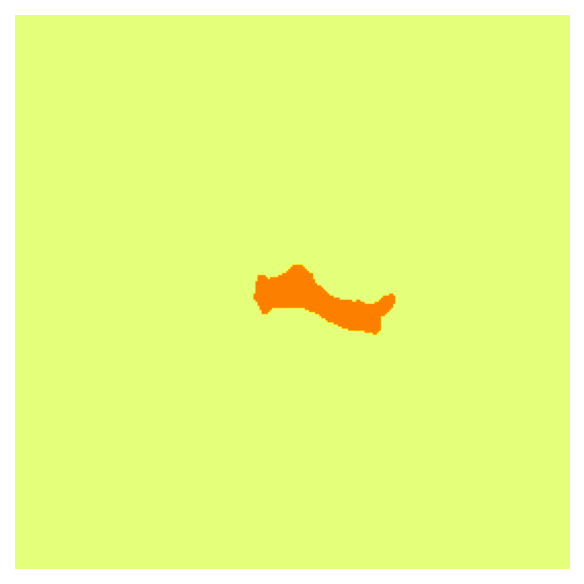} \\
    \includegraphics[width = 0.22 \linewidth]{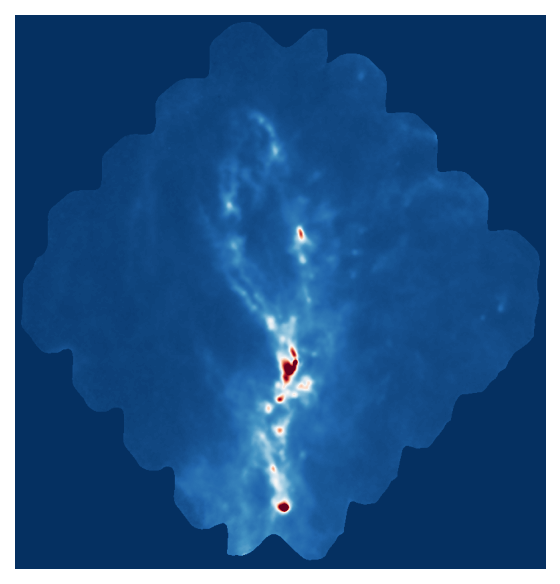} 
    \includegraphics[width = 0.22 \linewidth]{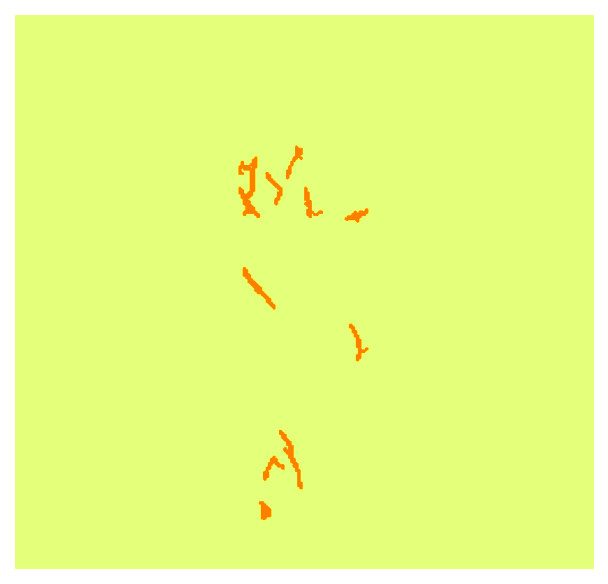}
    \hspace{0.3cm}
    \includegraphics[width = 0.22 \linewidth]{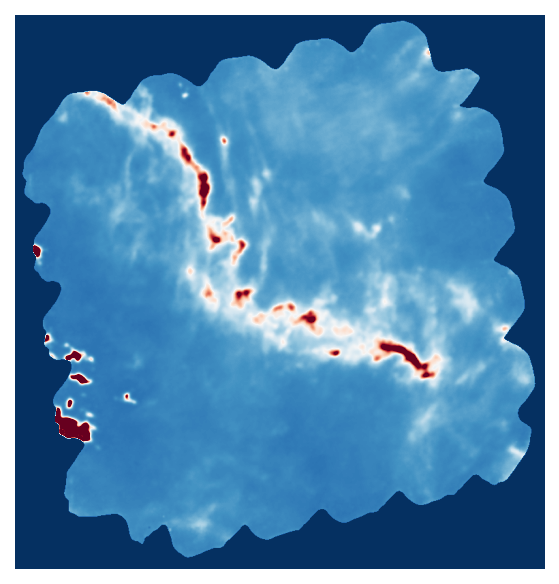}
    \includegraphics[width = 0.22 \linewidth]{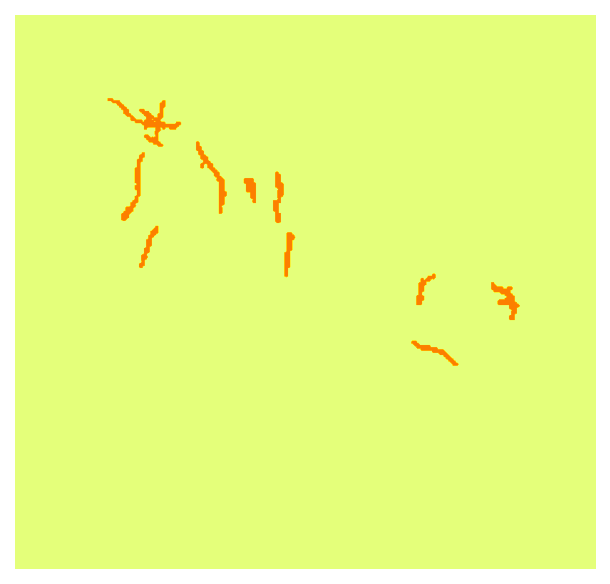}
    %\end{tabular}
    \caption{\quad Examples of maps in the training samples. First row: Cold Cores \textit{Planck}-based sample (Planck-cc), second row: all-sky \textit{Planck}-based sample (Planck-1), third row: HI4PI-based sample, fourth row: \textit{Herschel}-based sample.}
    \label{fig:examples}
\end{figure}

We ran each astronomical map through the RHT procedure and obtained the RHT intensity and angle. We then transformed RHT intensity maps to bitmap to obtain a mask. Thus, samples in our dataset contain three 2D maps of a particular region: an original intensity or column density map, a mask of identified filaments, and a map of filament angles. Examples are shown in Fig.~\ref{fig:examples}.

% == till here
\subsection{\textit{Planck}-based sample}
\label{sec:data_planck}
Planck filaments are represented in our sample in two ways.
The first Planck dataset is taken from the analysis of \cite{alina2019}, where RHT was applied over regions where Planck Galactic Cold Clumps are identified \citep{planck2016-XXVIII}. The Cold Clumps are regions that correspond to the coldest ISM, which are generally part of molecular clouds. The dataset contained {\nplanckcc} maps and the associated RHT outputs, such as maps of RHT intensity, angle, and angle uncertainty. The sub-sample will be denoted as "Planck-cc" in what follows. The angular size of the maps in this sub-sample was limited to two-by-two degrees, so we decided to complement it with yet another \textit{Planck}-based dataset from \cite{baimukhametova2021}.
The authors designed an algorithm dedicated to filament identification  and maps segmentation from large maps. It applies the RHT method on an arbitrarily chosen portion of the large map, labels all the detected structures, chooses the most prominent filament and determines the direction in which the map should be extended to capture the whole filament. As a result, it produces maps (mask and angle) that contain unique entire filaments while more minor features are masked. Finally, a visual analysis of the maps was performed to pick the most successful selections. The sub-sample will be denoted as "Planck-1". In total, the \textit{Planck}-based sample consists of {\nplancktot} maps.

\subsection{\textit{Herschel}-based sample}
We used \textit{Herschel} maps from the Galactic Cold Cores (GCC) survey \citep{juvela2010}, which consisted of 116 targets.
\textit{Herschel} telescope's angular resolution ($37"$) allows us to resolve the intriguing filamentary structures of molecular clouds. Thus, it diversifies our dataset and allows us to test the performance of neural networks at different complexity level because Planck filaments are generally smooth and extended because of the low angular resolution ($7'$ in our sample) of the telescope.
The corresponding column density maps were computed using spectral energy distribution fits with modified black-body law with a given spectral index $\beta = 2$ \citep{juvela2012}. The advantage of using the column density maps is having more prominent filaments and less marginal features, especially at low sensitivity observations, which is the case for \textit{Herschel}  compared to \textit{Planck}.

\textit{Herschel} GCC fields are an example of maps for which using the RHT method may be problematic because each map needs detailed parametrization. This is due to the variety of the shapes of the observed molecular clouds and their morphological complexity. To get the most robust training sample, we ran RHT with different parameters for each map and chose the output maps that gave the best MSSIM result. The kernel parameters, length, and width in pixels, are given in Table \ref{tab:kernels}. Their ranges extend from 1 pixel to 7 pixels for the width and from 5 to 31 pixels for the length. These values are motivated by the maps' size, and the variety of the structures observed in the maps.
Thus, each \textit{Herschel} map was treated with the most appropriate kernel that provided the mask and the angles map used in the neural network training. 
In general, a \textit{Herschel} GCC map contains many small structures because each observed structure is resolved and complex. Thus, after the RHT procedure, we additionally perform mask binarization and select only ten largest filaments with respect to their pixel count. This choice is motivated by the necessity to avoid noisy input and non-significant structures, improving generalization. 

\begin{table}[h]
    \caption{ Kernels' length ($l$) and width ($w$) in pixels, used in the RHT procedure to generate training sample based on \textit{Herschel} GCC maps}
    \centering
    \begin{tabular}{c|c|c|c|c|c|c|c|c}
         \diagbox{$w$}{$l$} &   5&  7&  9&  13& 15& 21& 27  &31  \\ \hline \hline
         1 & \checkmark &   & \checkmark  &  \checkmark &   &   &   & \\ \hline
         3 &  & \checkmark  &   &   & \checkmark  & \checkmark  & \checkmark  & \checkmark \\ \hline
         5 &  &   &   &   & \checkmark  &   &   & \checkmark
    \end{tabular}
    
    \label{tab:kernels}
\end{table}

\subsection{HI4PI-based sample}
 We used the all-sky survey of the atomic hydrogen data from the Effelsberg-Bonn and Parkes telescopes which is publicly available as the HI4PI survey \citep{benbekhti2016}. The atomic hydrogen (HI) traces the diffuse gas content and is generally mixed with dust in the ISM. In addition, findings of \citep{boulanger1996,kalberla2016} showed that the HI gas and \textit{Planck} dust filaments show good agreement. Thus, HI data adds another dimension to the angular scales in our dataset, with $16'$ resolution. We ran the RHT procedure through the column density maps of HI at the same positions where \textit{Planck} filaments were found in the study by \cite{baimukhametova2021}. We conducted a visual check-up and obtained {\nhi} maps.

\section{Results}
\label{sec:results}
In this Section, we present the results of application of two most efficient architectures for our purposes, although other models were tested and will be discussed in Section \ref{sec:discussion}. The first architecture is based on Mask R-CNN model. It is used to identify the location and the shape of the filaments, that is, to produce segmentation masks. The masks give the location and extent of the identified filaments. The second architecture is based on the U-Net model. It is used to determine the filaments orientation angles from the mask.

\subsection{Mask R-CNN learning model}
% talk with Adai about the paragraph!
In our study, we use Mask-RCNN network architecture \cite{he2017} to extract masks of filament structures. 
Mask R-CNN is a Convolutional Neural Network (CNN) and state-of-the-art in terms of image segmentation and instance segmentation.
It generates bounding boxes and segmentation masks for each instance of an object in the image. 
In other words, it detects each object in an image and provides information about its position. 
Mask R-CNN is based on ResNet101 backbone and Feature Pyramid Network (FPN).
%% Adai: refer to the figure and add clear text

\subsubsection{Training}
In this work, we decided to train the neural network using training sets of astronomical images and their masks generated by means of the RHT procedure. 
The training datasets consist of astronomical maps described in Section \ref{sec:data} and their corresponding masks. The latter are obtained from RHT intensity maps to which we applied binary thresholding.
The final objective of the neural network was to identify filamentary structures. All images were resized to $256 \times 256 \times 3$ pixels dimension.
Technically, the Mask R-CNN neural network was implemented based on Keras \cite{chollet2015keras} and Tensorflow libraries \cite{tensorflow2015-whitepaper} with the support of the GPU assisted parallel computations. We trained the network for 30 epochs. To improve the process, we applied transfer learning which was enabled based on the COCO data set \cite{cocodataset}. 

%During the construction of the neural network, we found the "from simple to complex" approach to be inefficient. 
We first ran the network on the Planck-1 data set where each map has only one identified filament. 
However, experimental results showed the critical importance of feeding the neural network with multiple filaments in the target image during the training phase. If the neural network was provided with a single filament for each map at the training stage, as in the case of Planck-1 sub-sample, the network could not detect many significant structures in the test set. This explains low mean Average Precision with increasing training size in Fig.~\ref{fig:map_rcnn}. 
Once the criterion of a single filament is relaxed, there is no critical difference related to the number of input filaments, as we see in the \textit{Herschel}-based data set. However, increasing the number of significant filaments led to detection of noisy structures that blended the results.

\subsubsection{Results}

\begin{figure}[ht]
    \centering
   \begin{tabular}{ccc}
   Original image & Mask R-CNN & RHT \\
   \small{} & \small{0.3001} & \small{0.2981} \\ %G82.65-2.00
   \includegraphics[height = 0.3 \linewidth]{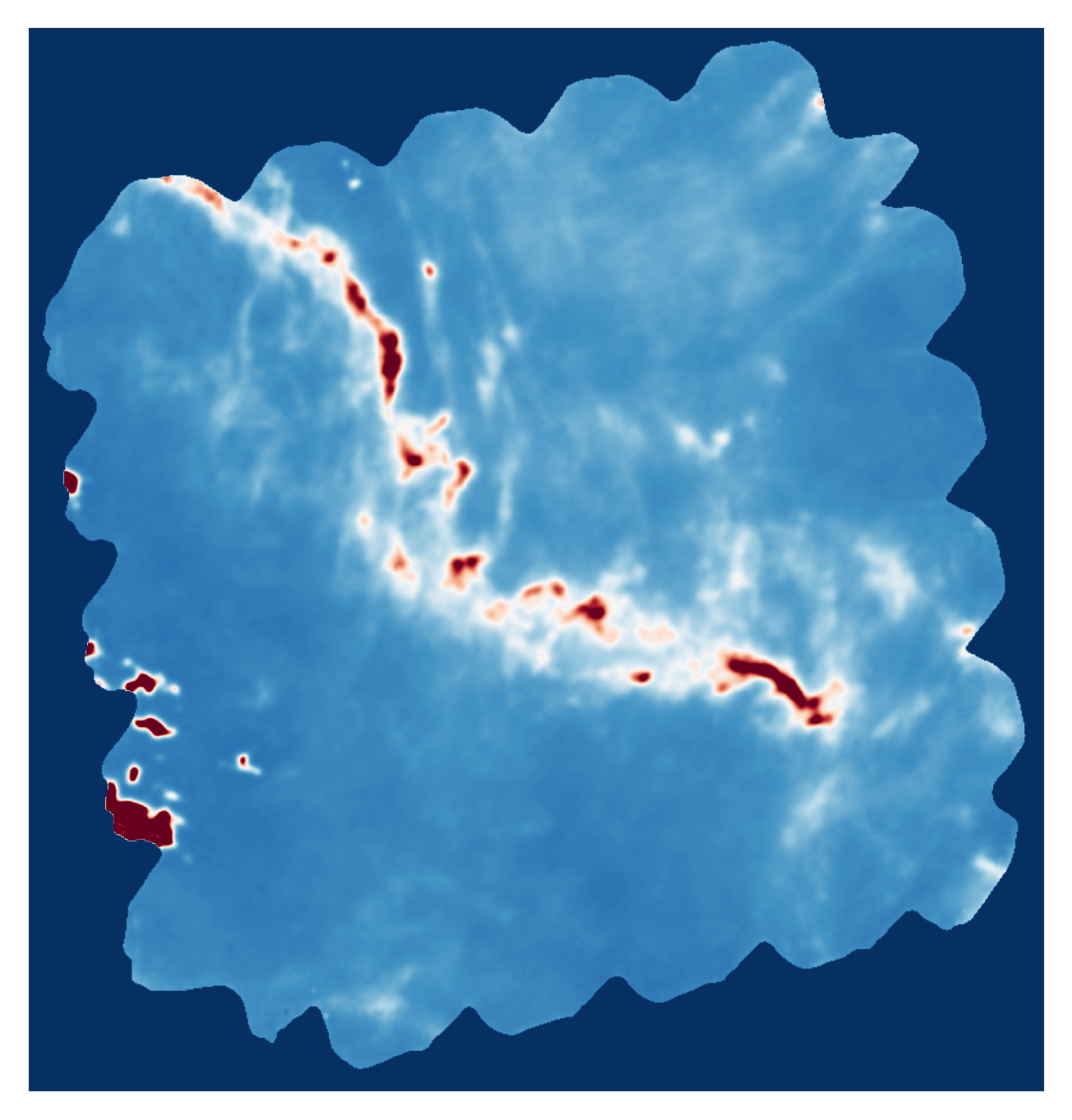} 
   &
   \includegraphics[height = 0.3 \linewidth]{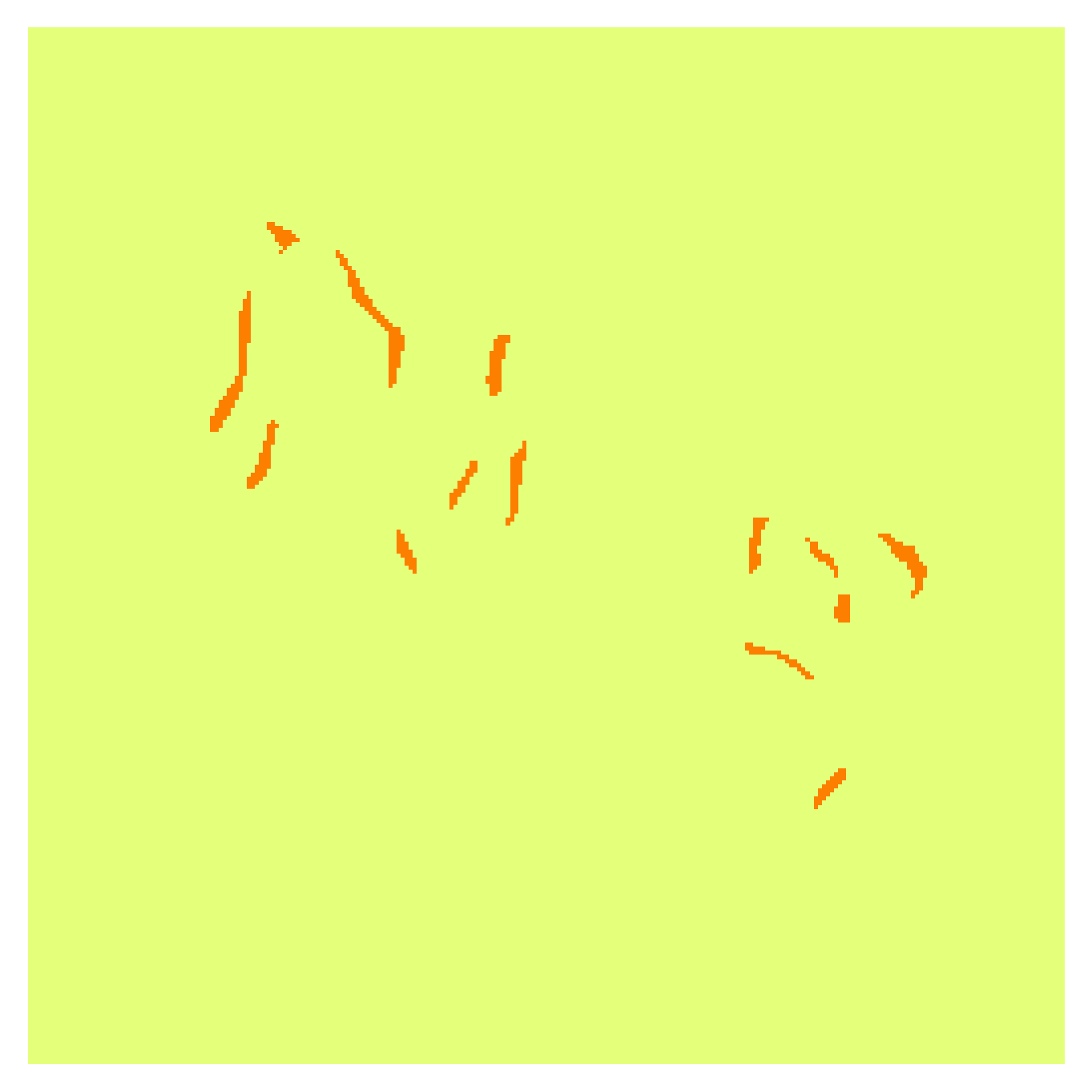}
   &
   \includegraphics[height = 0.3 \linewidth]{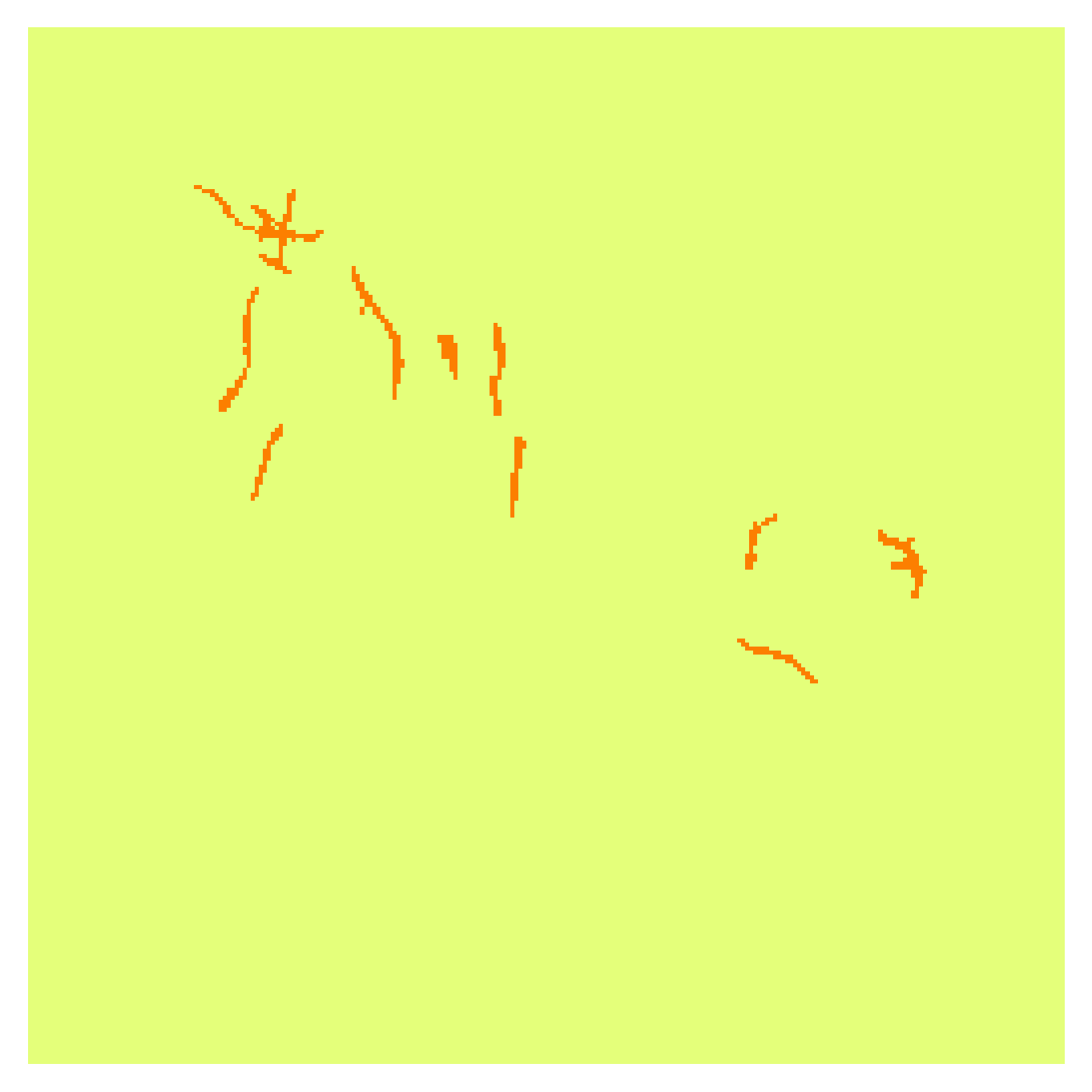} \\
%    \small{} & \small{0.5298} & \small{0.5307} \\ %G202.02+2.85
%\includegraphics[height = 0.3 \linewidth ]{Figures/v7_hs_G202.02+2.85.png} &
% \includegraphics[height = 0.3 \linewidth ]{Figures/v7_hs_G202.02+2.85_maskrcnn.png} &
% \includegraphics[height = 0.3 \linewidth ]{Figures/v7_hs_G202.02+2.85_mask10.png} \\
\small{} & \small{0.2881} & \small{0.2885} \\ %G210.90-36.55
\includegraphics[height = 0.3 \linewidth]{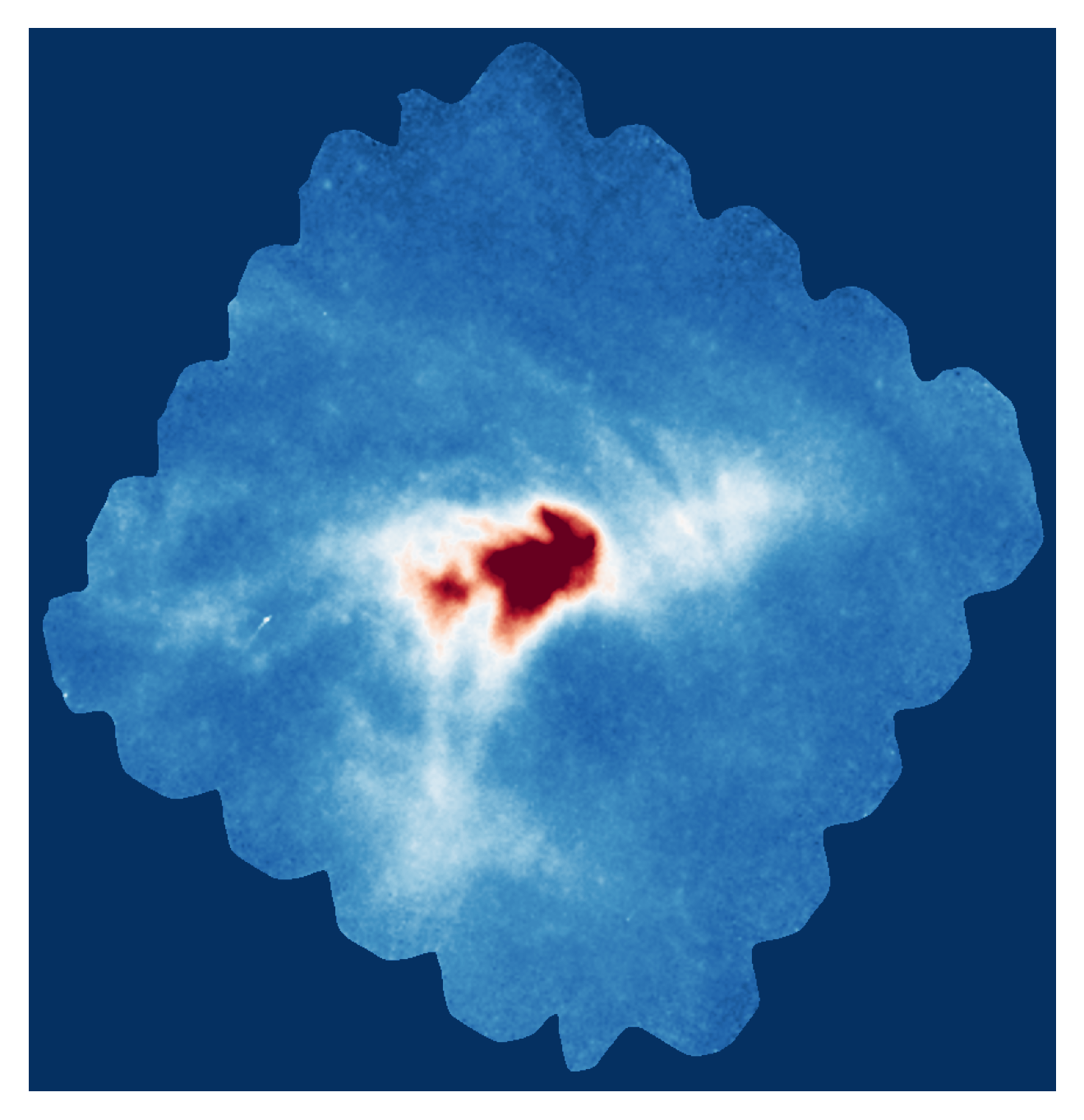}
&
\includegraphics[height = 0.3 \linewidth]{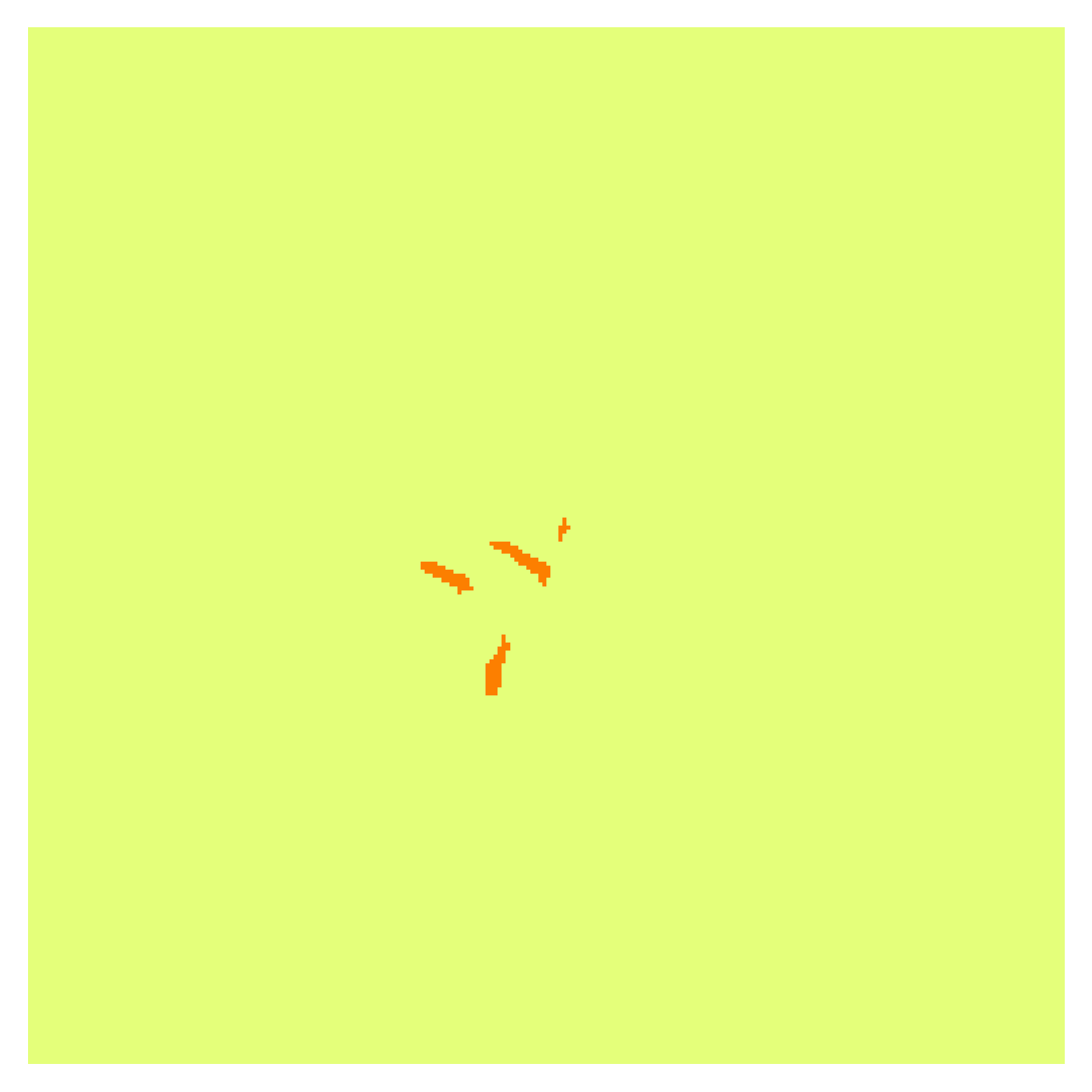}
&
\includegraphics[height = 0.3 \linewidth]{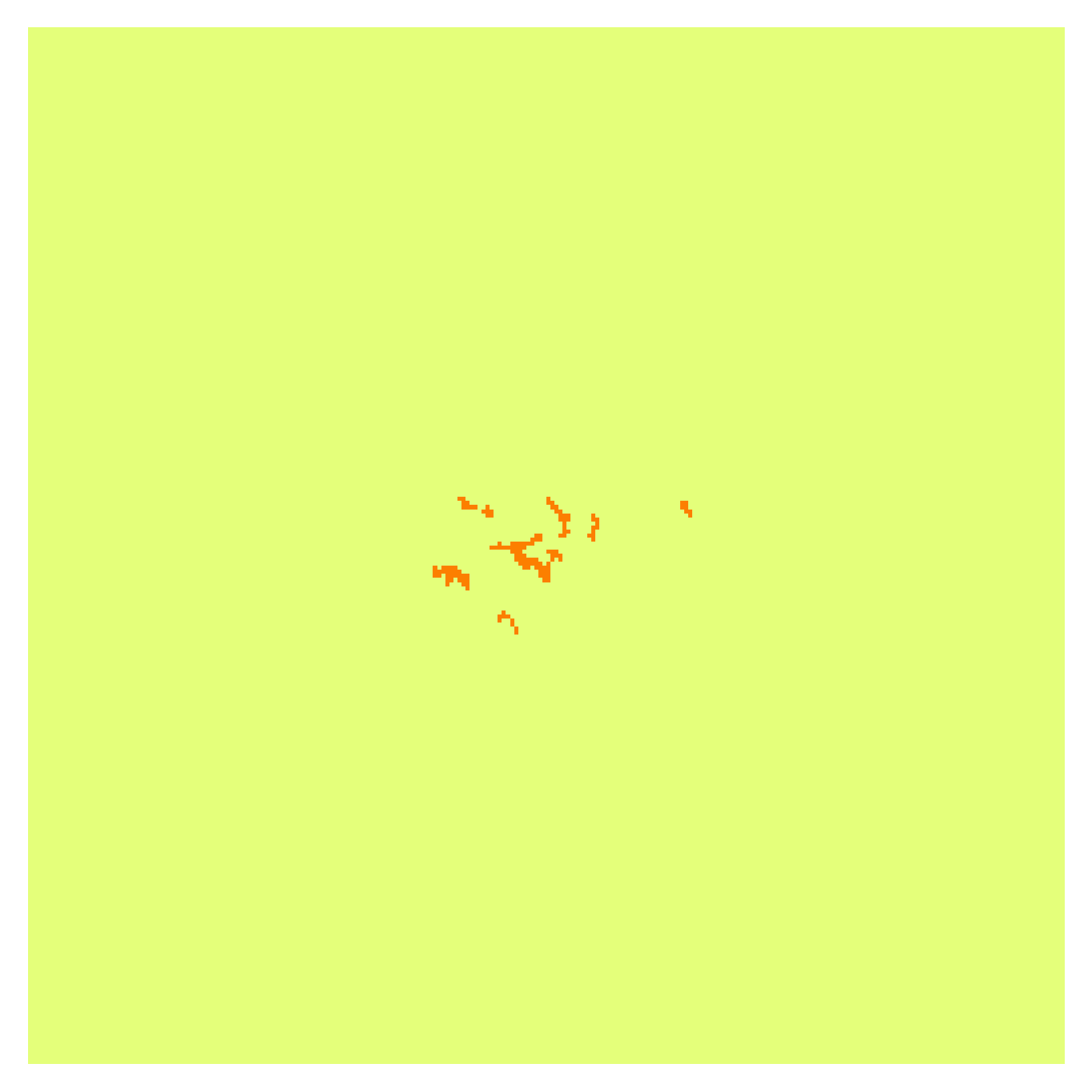}
\\
\small{} & \small{0.3048} & \small{0.3040} \\
\includegraphics[height = 0.3 \linewidth]{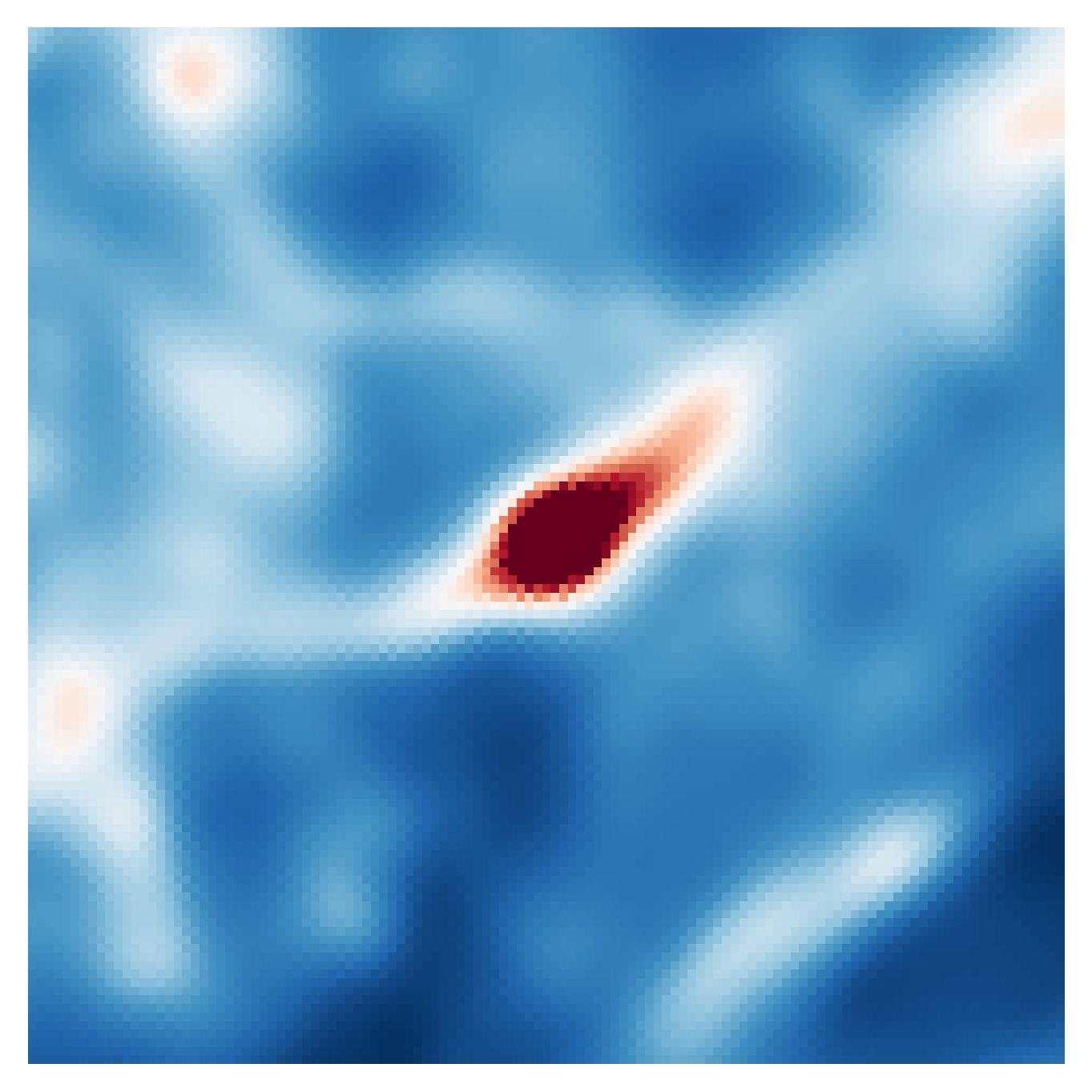}
&
\includegraphics[height = 0.3 \linewidth]{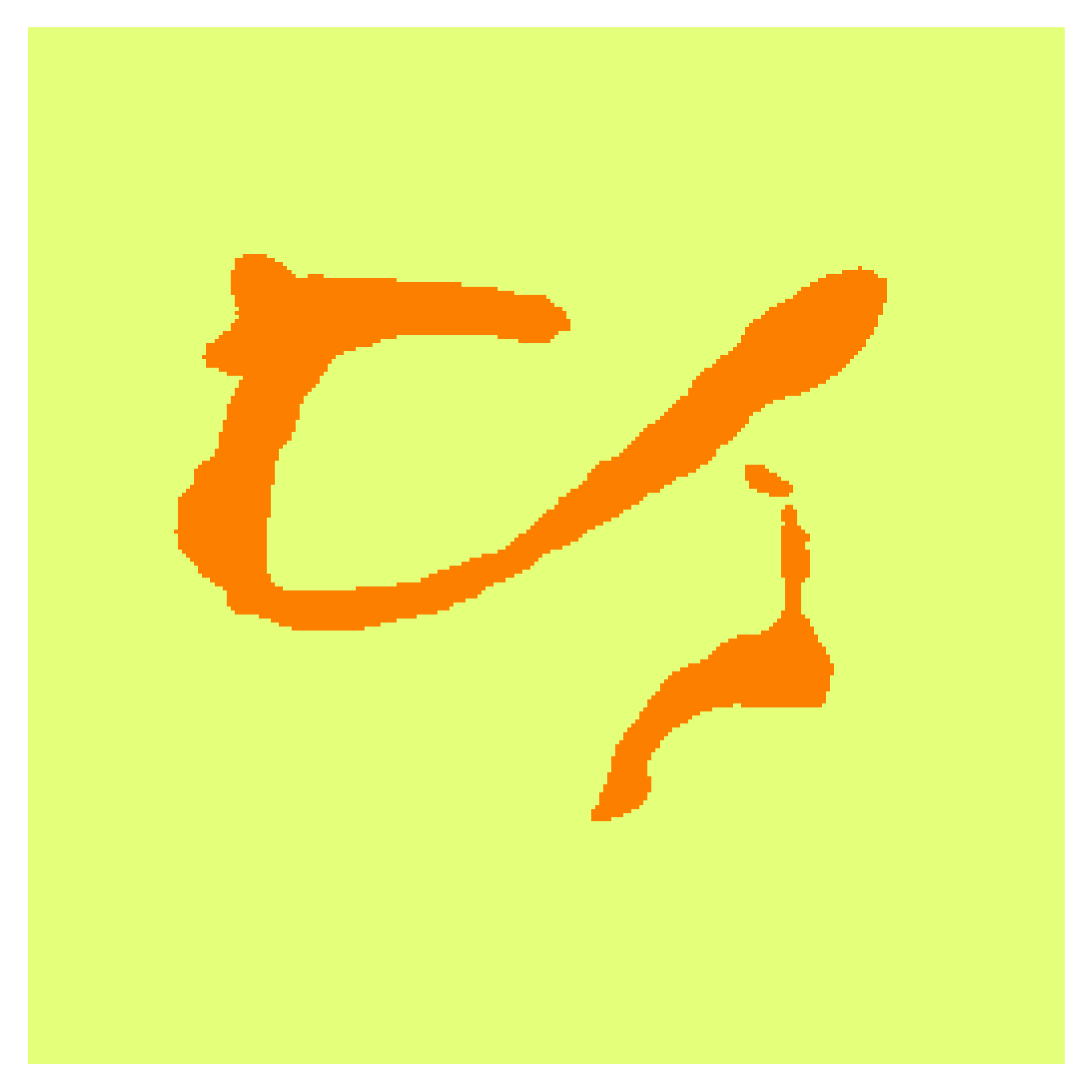}
&
\includegraphics[height = 0.3 \linewidth]{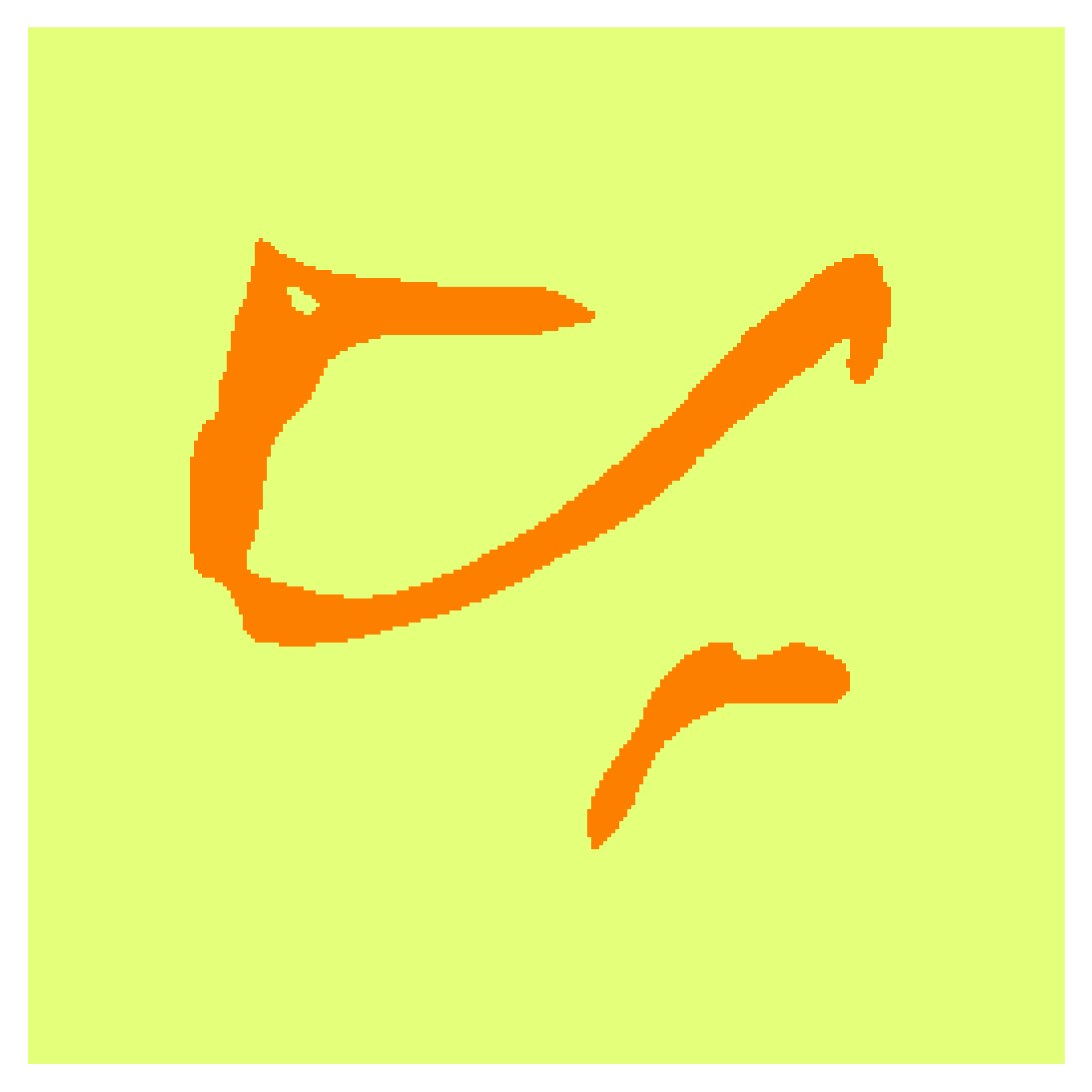}
\\
\small{} & \small{0.1904} & \small{0.1910} \\
\includegraphics[height = 0.3 \linewidth]{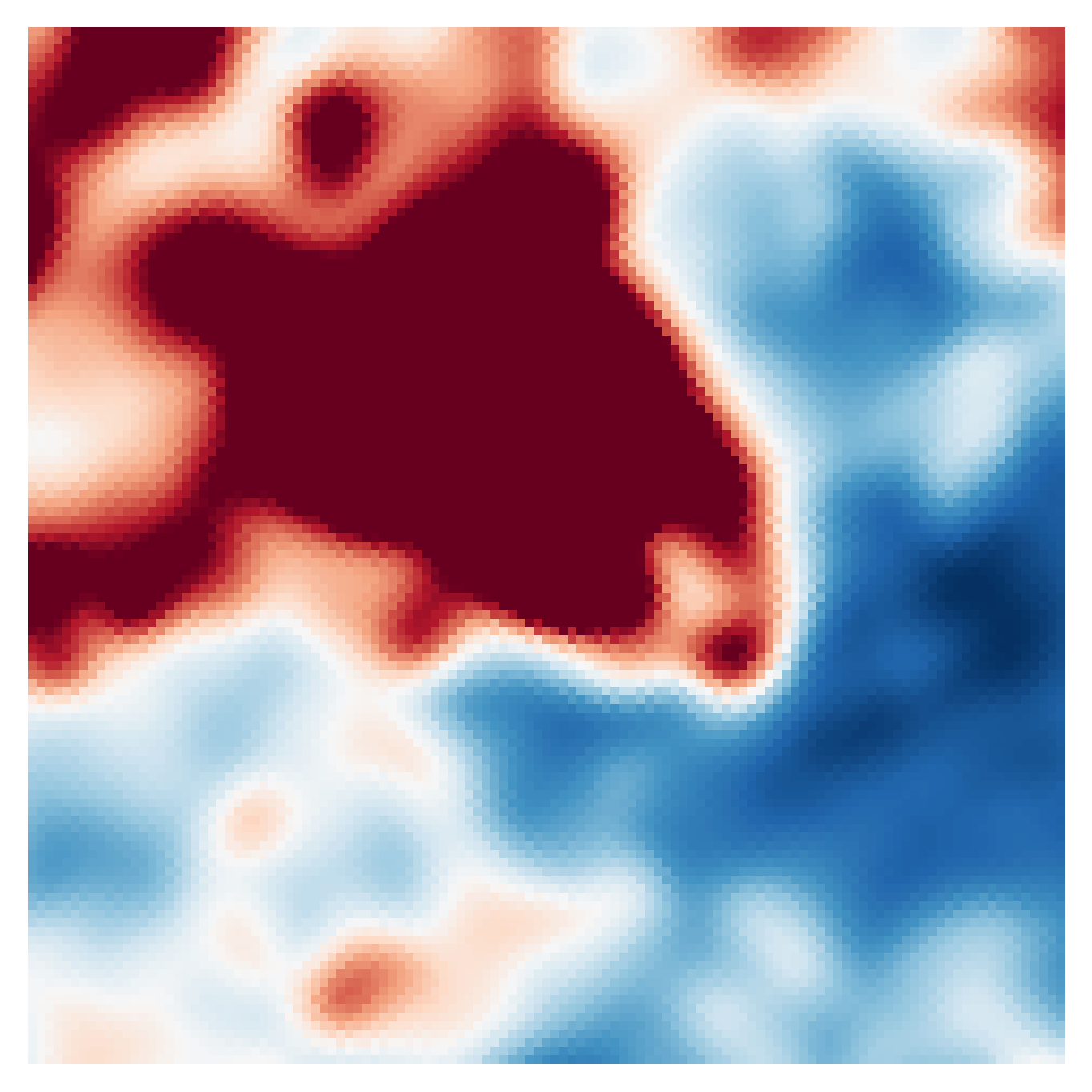}
&
\includegraphics[height = 0.3 \linewidth]{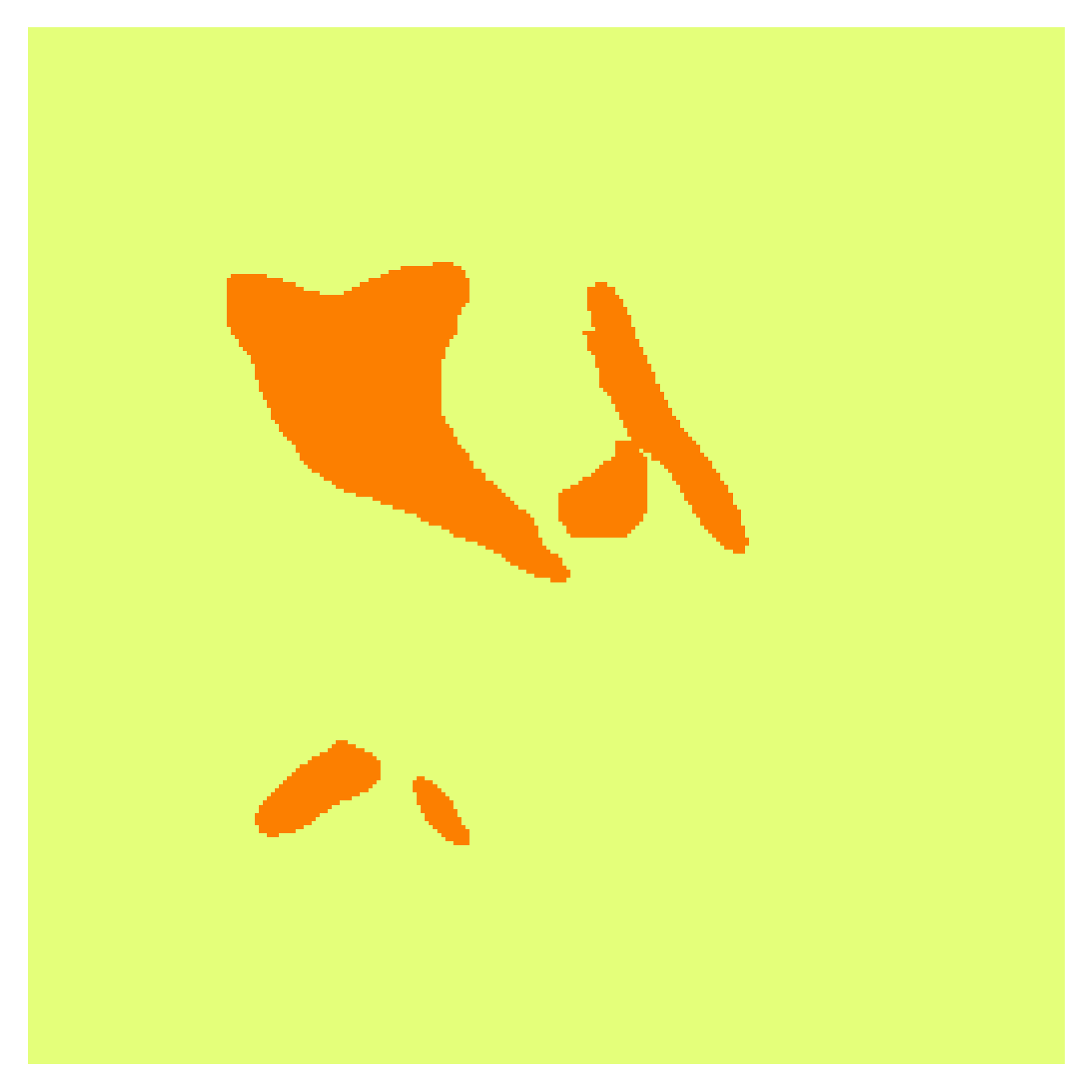}
&
\includegraphics[height = 0.3 \linewidth]{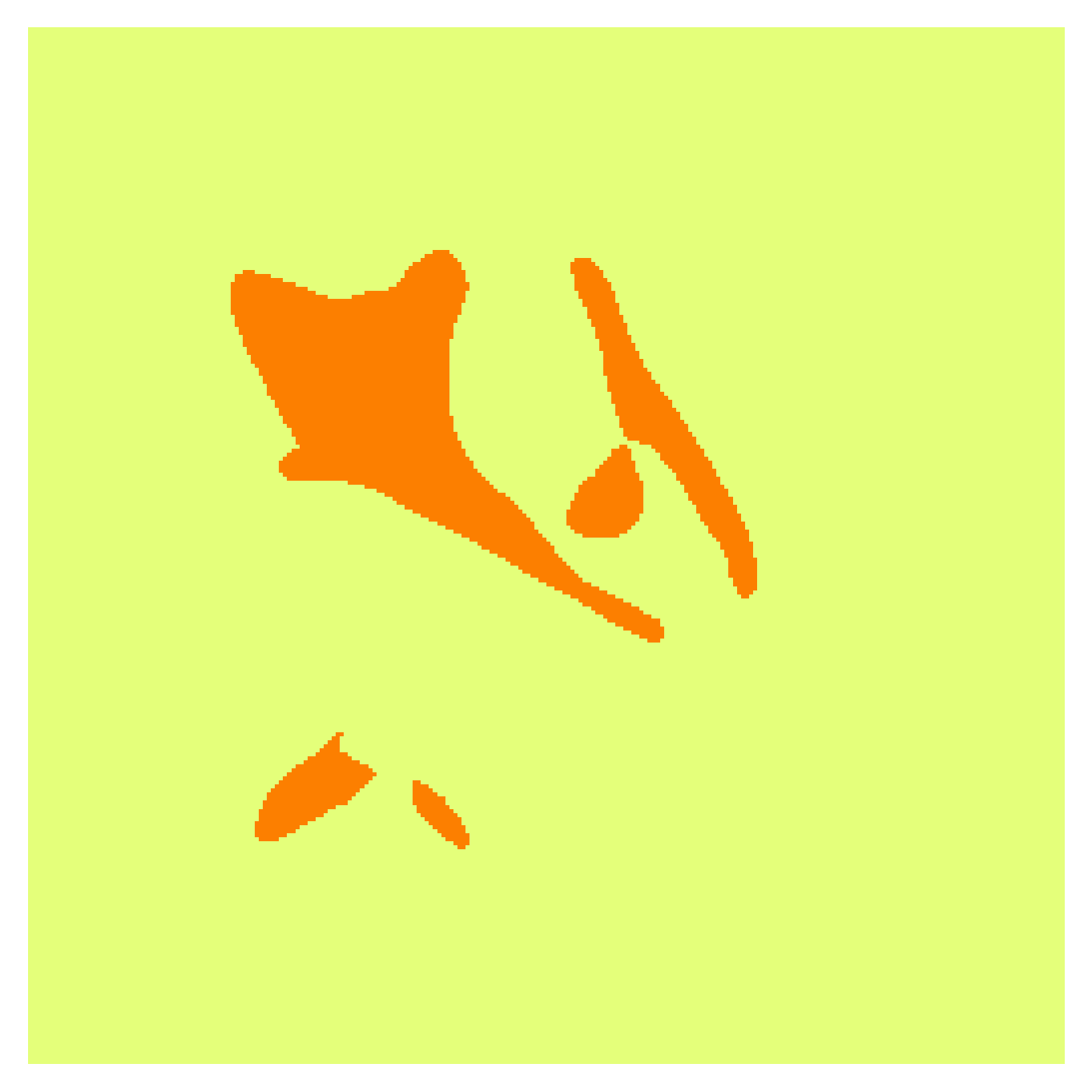} \\
\small{} & \small{0.3127} & \small{0.3086} \\
\includegraphics[height = 0.3 \linewidth]{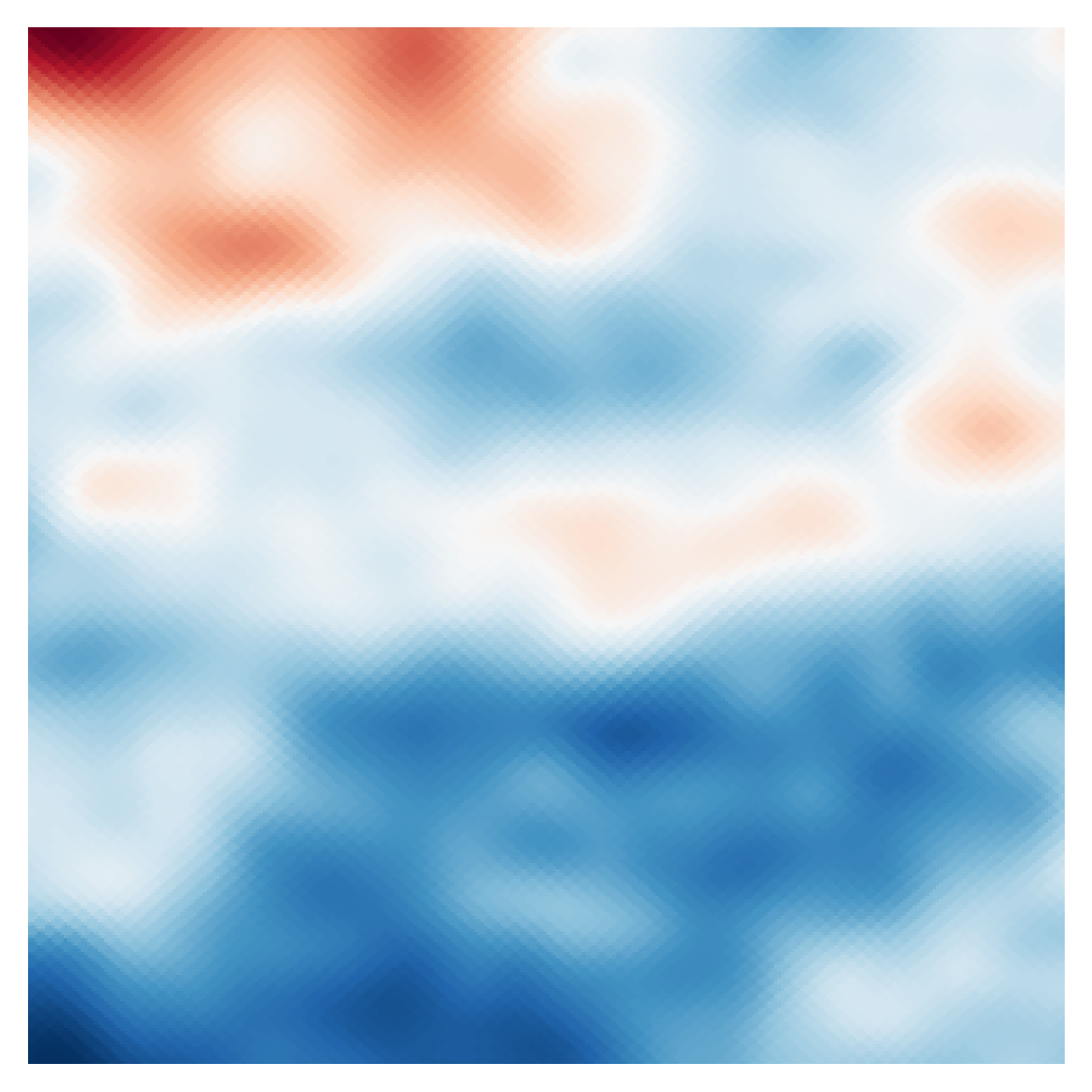}
&
\includegraphics[height = 0.3 \linewidth]{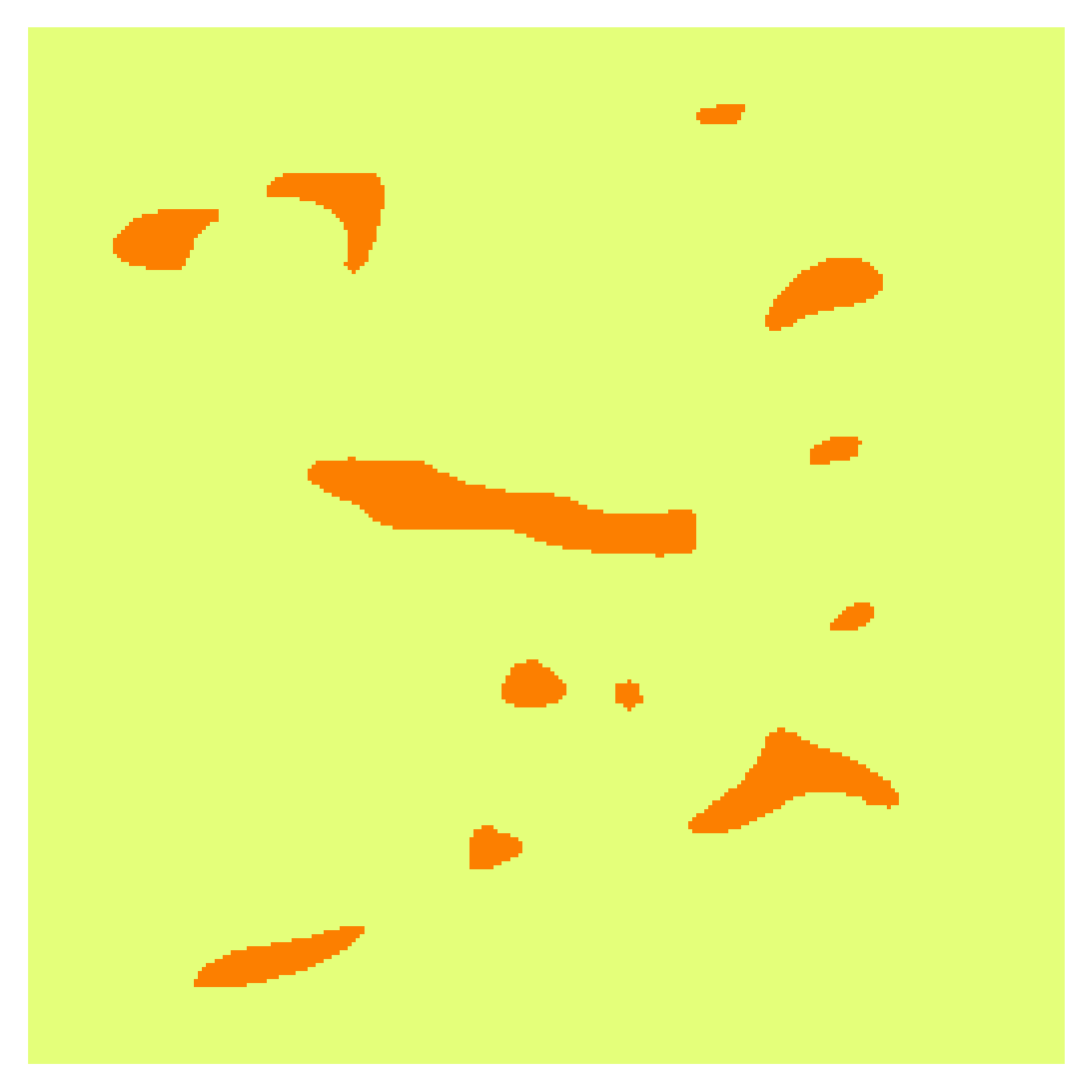}
&
\includegraphics[height = 0.3 \linewidth]{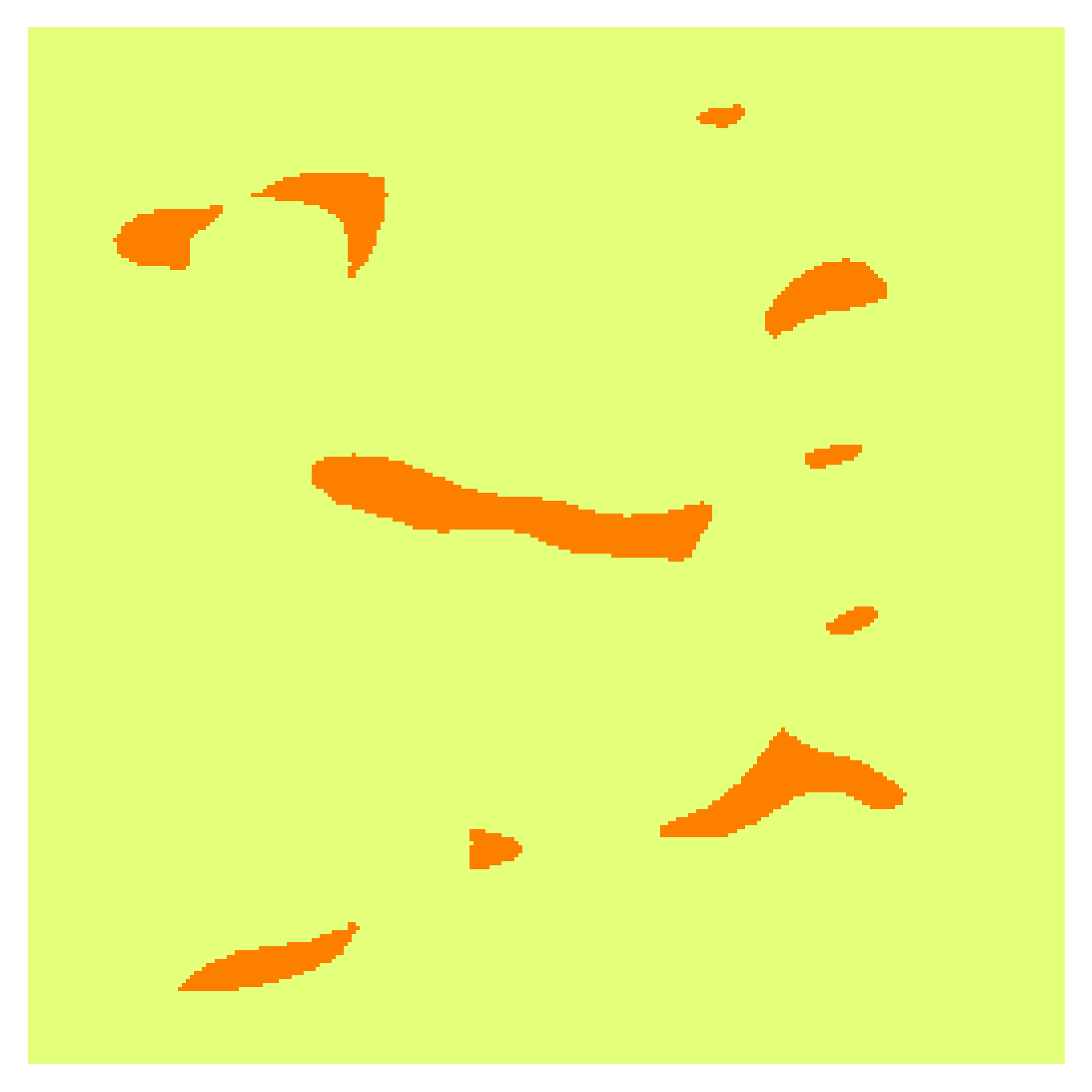}
\end{tabular}

    \caption{\quad Examples of obtained filaments: the first column depicts the original intensity maps, the second column shows results of the Mask R-CNN and the last column shows results of the RHT procedure. Numbers in the second and third columns indicate the structural similarity score measured by the MSSIM metric.}
    \label{fig:rcnn_examples}
\end{figure}

\begin{figure}[ht]
    \centering
    \includegraphics[width = \linewidth]{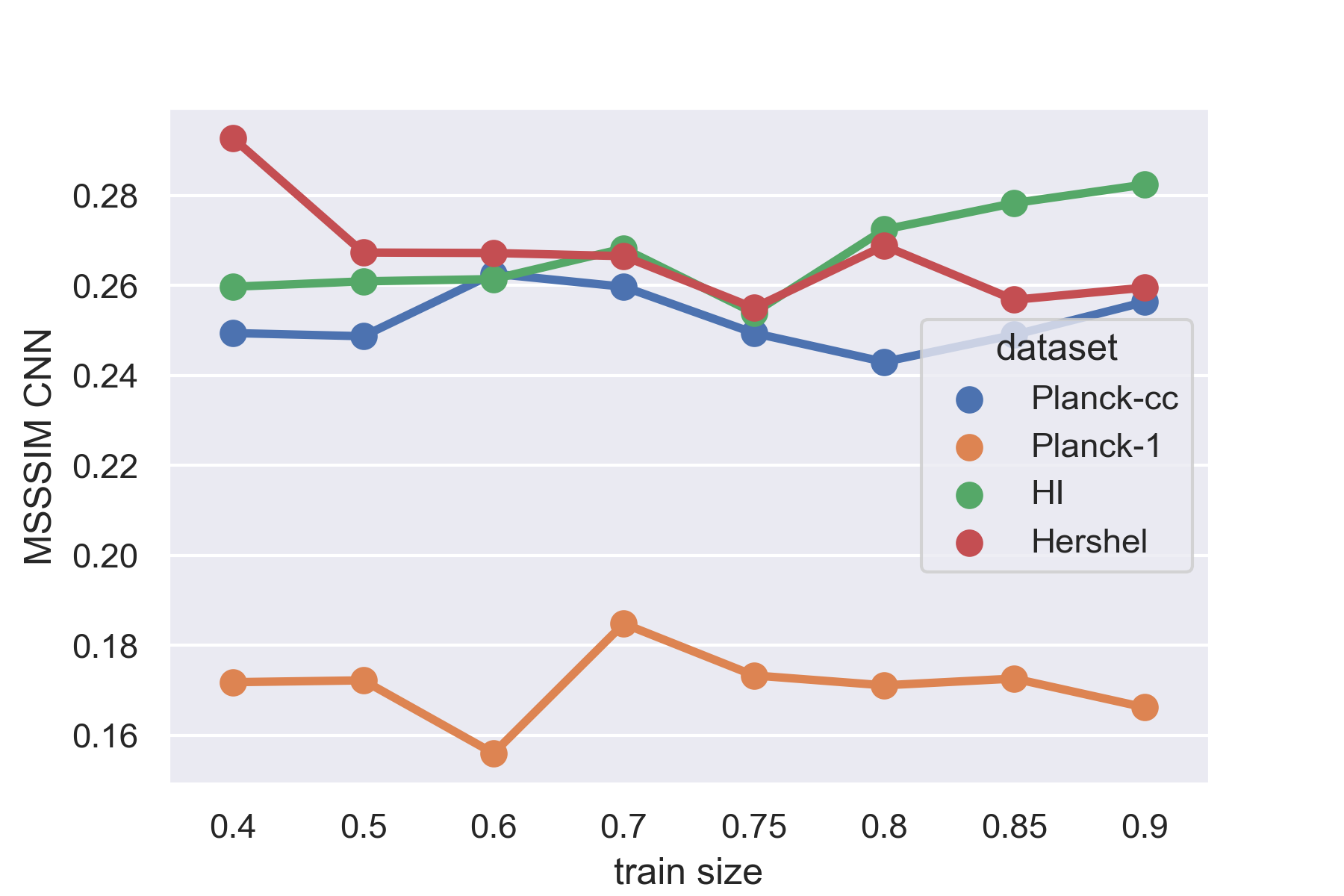} 
    \caption{\quad MSSIM values over the different training sample sizes for the Mask R-CNN neural network.}
    \label{fig:mssim_rcnn}
\end{figure}

\begin{figure}[ht]
    \centering
    \includegraphics[width = \linewidth]{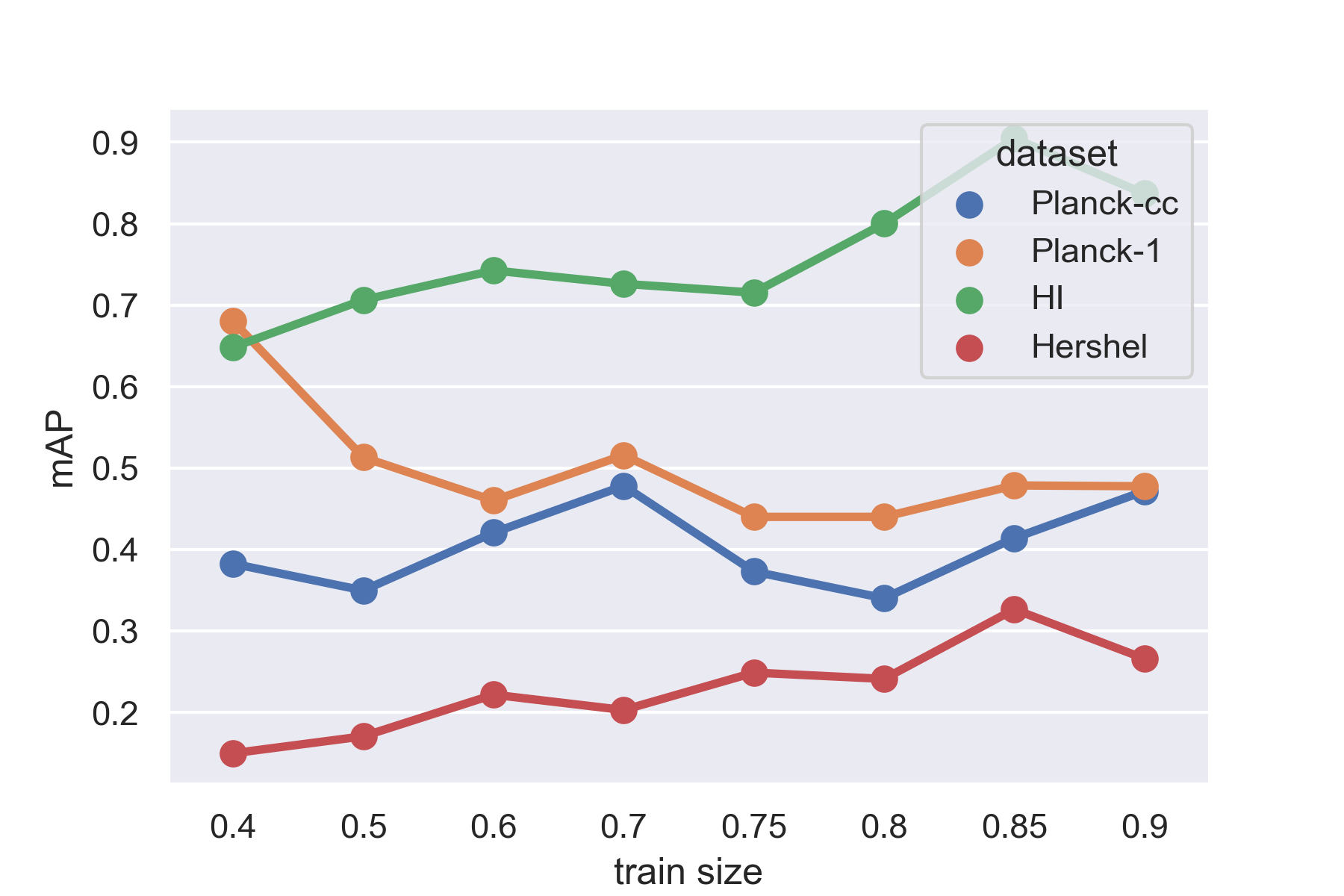} 
    \caption{\quad Mean Average Precision (mAP) values over the different training sample sizes, normalised to one with respect to the total number, for the Mask R-CNN neural network.}
    \label{fig:map_rcnn}
\end{figure}

We trained the neural network separately on each of the available data sets.

We show in the central column of Fig.~\ref{fig:rcnn_examples} sample outputs of the Mask R-CNN, and the corresponding RHT masks in the left column. Qualitatively, Mask R-CNN yields larger and smoother structures than the RHT procedure and detects the most significant structures.
%, although quantitatively, the MSSIM value may be larger for the RHT output. In general, the MSSIM value is larger for the neural network in $ n \%$ of the cases depending on the dataset.
Quantitatively, we assess the performance of the neural network compared to RHT using the MSSIM index and the mean average precision score. The MSSIM index, described in Section \ref{sec:methods}, quantifies how the output masks differ in structure compared to the original map. A high score correlates with a high similarity between entities and a low score informs about less significant similarities. The score ranges from -1 to 1. %We also compute the mean average precision value.
We also assess the results using the mean average precision (mAP).
The mean average precision shows the average precision score
for the fixed value of intersection over union (IOU) of 0.5, i.e. the proportions of
test samples that has higher score than the predefined IOU score.
The average precision is computed according to the following equation:
\begin{equation}
    \text{AP} = \sum_n (R_n - R_{n-1}) P_n   \, , 
\end{equation}
where $R_n$ and $P_n$ depict the recall and precision values at n-th threshold. The mean average precision is a mean value of AP across all data set. 

We summarize in Table \ref{tab:maskrcnn} the performance of the Mask-RCNN model by calculating the mAP score and the mean and standard deviation of the differences between MSSIM indexes. On average, in terms of the morphological similarity of the output, neural network performs at least as well or better than the RHT.

%Visual inspection of samples obtained from our proposed method as well as the quantification using mAP show that this algorithm performs better than RHT method in some cases \textbf{some cases? how many cases?}. 
%Also, the MSSIM value is larger for the neural network in $ n \%$ of the cases depending on the dataset.

Fig.~\ref{fig:mssim_rcnn} shows the comparison between the RHT generated filament masks and the output masks produced by the neural network based on MSSIM index. For the HI filaments, the larger the training size, the higher the MSSIM value, meaning that the the neural network becomes more efficient. As for the \textit{Planck}- and \textit{Herschel}-based data sets, the morphological similarity comparison does not show any clear trend with increasing size. We note that the Planck-1 sub-sample shows low performance because maps contain only one, largest filament, by construction. 

Fig.~\ref{fig:map_rcnn} shows the accuracy of the filament identification measured in terms of mean average precision score depending on the size of the training sample, normalized to one. Neural network trained on \textit{Herschel}-based and HI4PI-based data sets shows a steady increase in accuracy as the sample size is increased while \textit{Planck}-based sub-samples show more peculiar behaviors. The Planck-cc sub-sample does not show significant improvements, while the Planck-1 sub-sample shows an evident lack of efficiency that correlates with the size of the training sample.

\renewcommand{\arraystretch}{2.0}
\begin{table*}
\caption{Estimation of performance of the Mask R-CNN training. The mean Average Precision (mAP) score measures the overlap between the neural network result and the RHT result. Difference between the MSSIM from Mask R-CNN and RHT results measures the efficiency of the neural network over RHT regarding morphological structure compared to the original image}
\setlength{\tabcolsep}{7pt}
\begin{tabular}{llllll}
\hline
\multicolumn{1}{|l|}{dataset} & \multicolumn{1}{|p{1.6cm}|}{\centering train set \\sample size} & \multicolumn{1}{|p{1.6cm}|}{\centering validation set \\ sample size} & \multicolumn{1}{l|}{mAP score} & \multicolumn{1}{l|}{mean(MSSIM$_{\mathrm{NN}}$ - MSSIM$_{\mathrm{RHT}}$)} & \multicolumn{1}{l|}{std(MSSIM$_{\mathrm{NN}}$ - MSSIM$_{\mathrm{RHT}}$)} \\ \hline
Hershel-based                            &  90                                                & 24                                                       & 0.32                                & 0.07                                & 0.03                                 \\
HI4PI-based                 &  120                                                & 28                                                      &  0.35                              &   0.08                              & 0.04 \\
Planck-cc                 &  100                                                & 37                                                      &  0.33                              &   0.03                              & 0.0035 \\
Planck-1 & 85 & 20 & 0.37 & 0.037 & 0.004
\end{tabular}
 
    \label{tab:maskrcnn}
\end{table*} 

\subsection{U-Net model}
For estimating the orientation angle of the filaments, we propose the U-Net deep learning architecture \cite{ronneberger2015u}. The U-Net model was originally used for biomedical image segmentation, but later was adopted for various domains and proved to be effective due to combination of local and global features in the process of making the resulting prediction. 

\begin{figure}
    \centering
   \begin{tabular}{ c c c }
   Original image & Unet & RHT\\
   \small{} & \small{0.3001} & \small{0.2981} \\ %G82.65-2.00
   \includegraphics[height = 0.27 \linewidth]{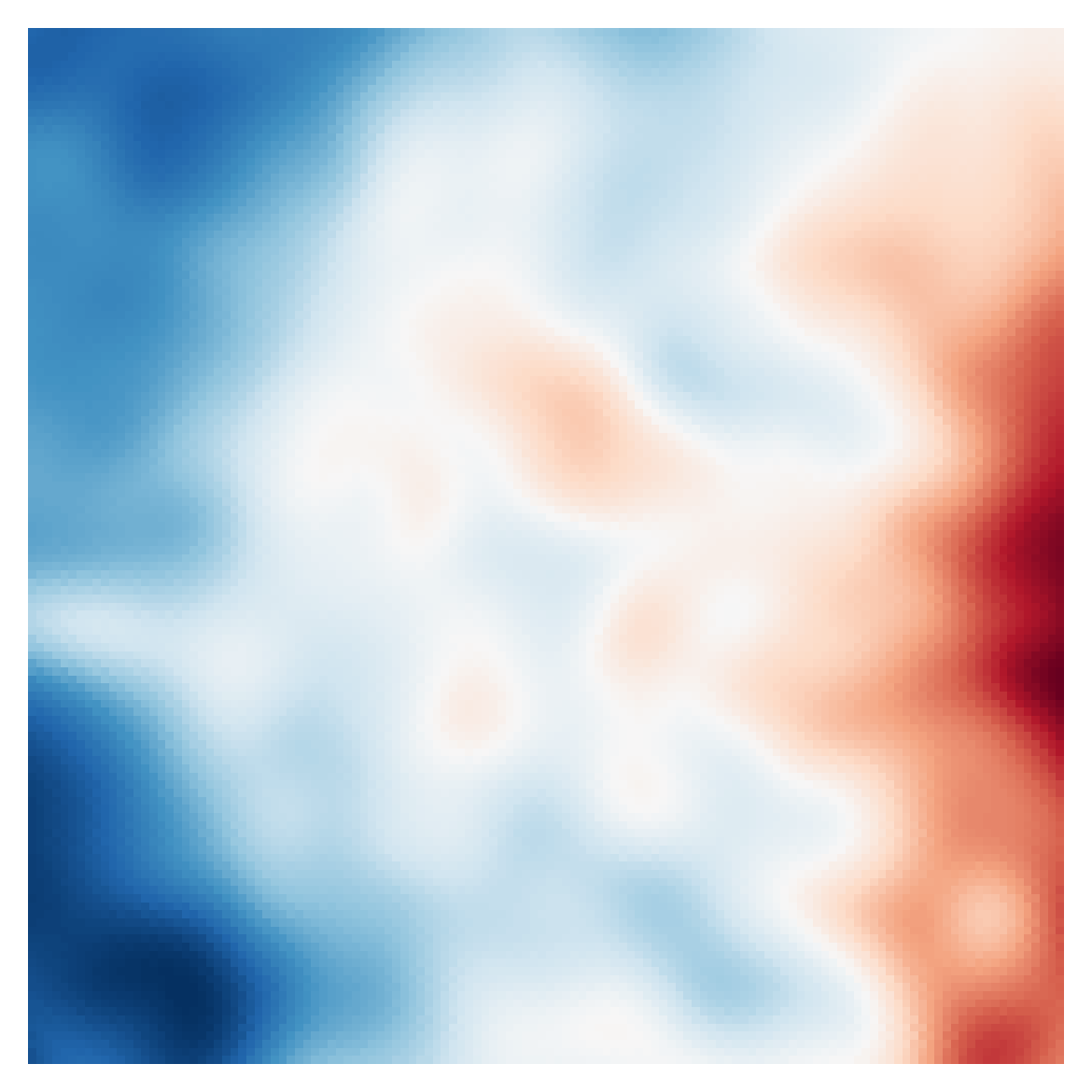} 
   &
   \includegraphics[height = 0.27 \linewidth]{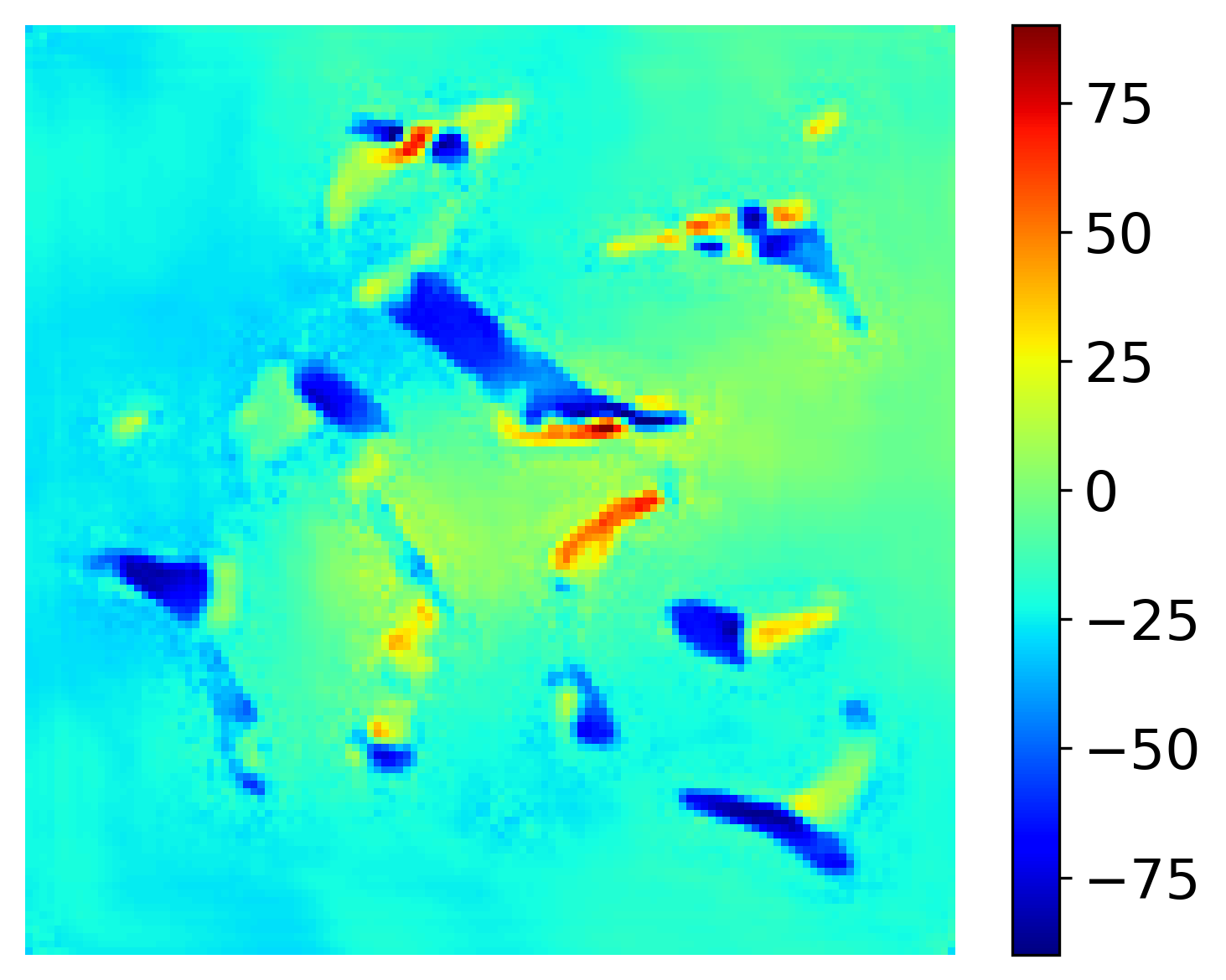}
   &
   \includegraphics[height = 0.27 \linewidth]{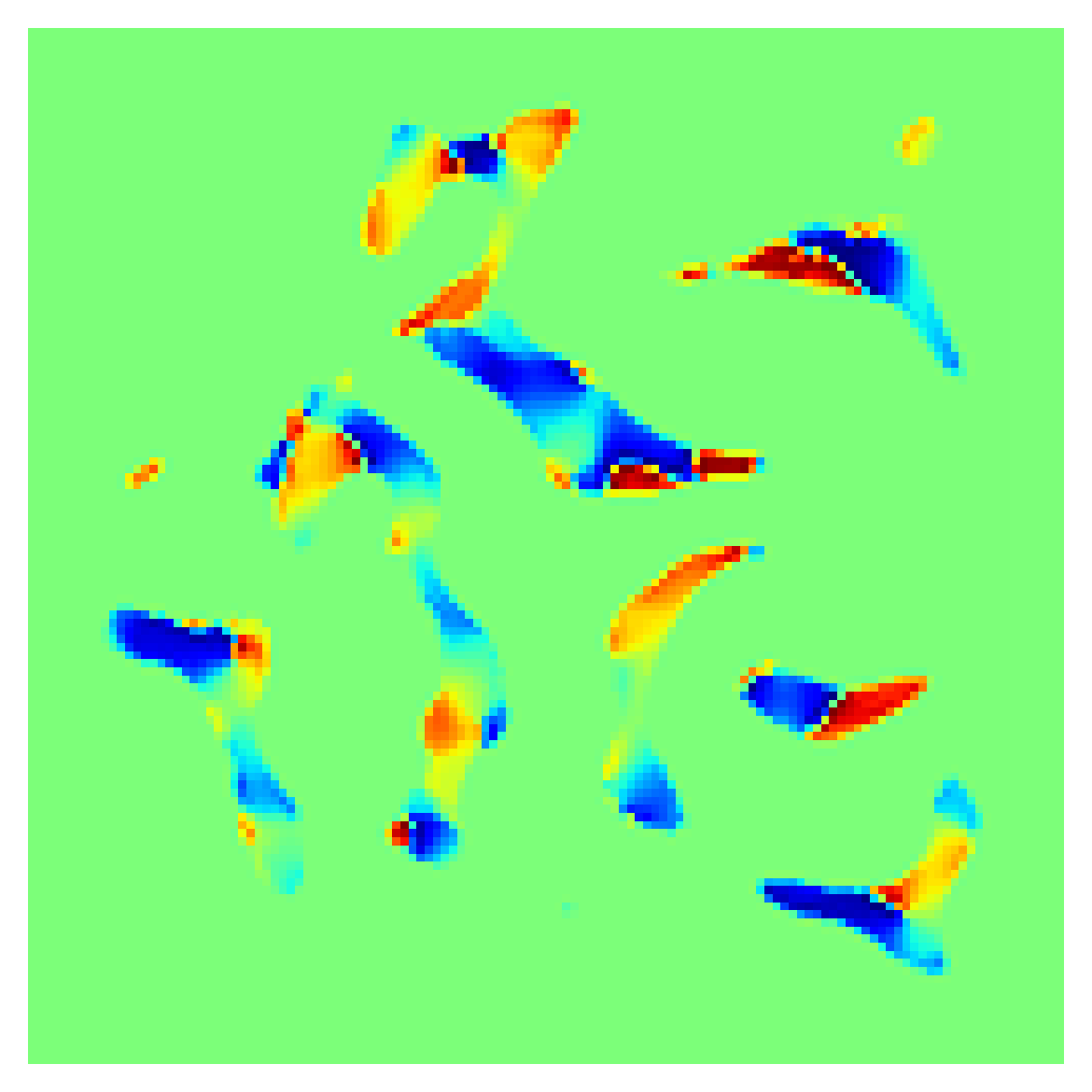} \\
   \small{} & \small{0.3001} & \small{0.2981} \\ %G82.65-2.00
   \includegraphics[height = 0.27 \linewidth]{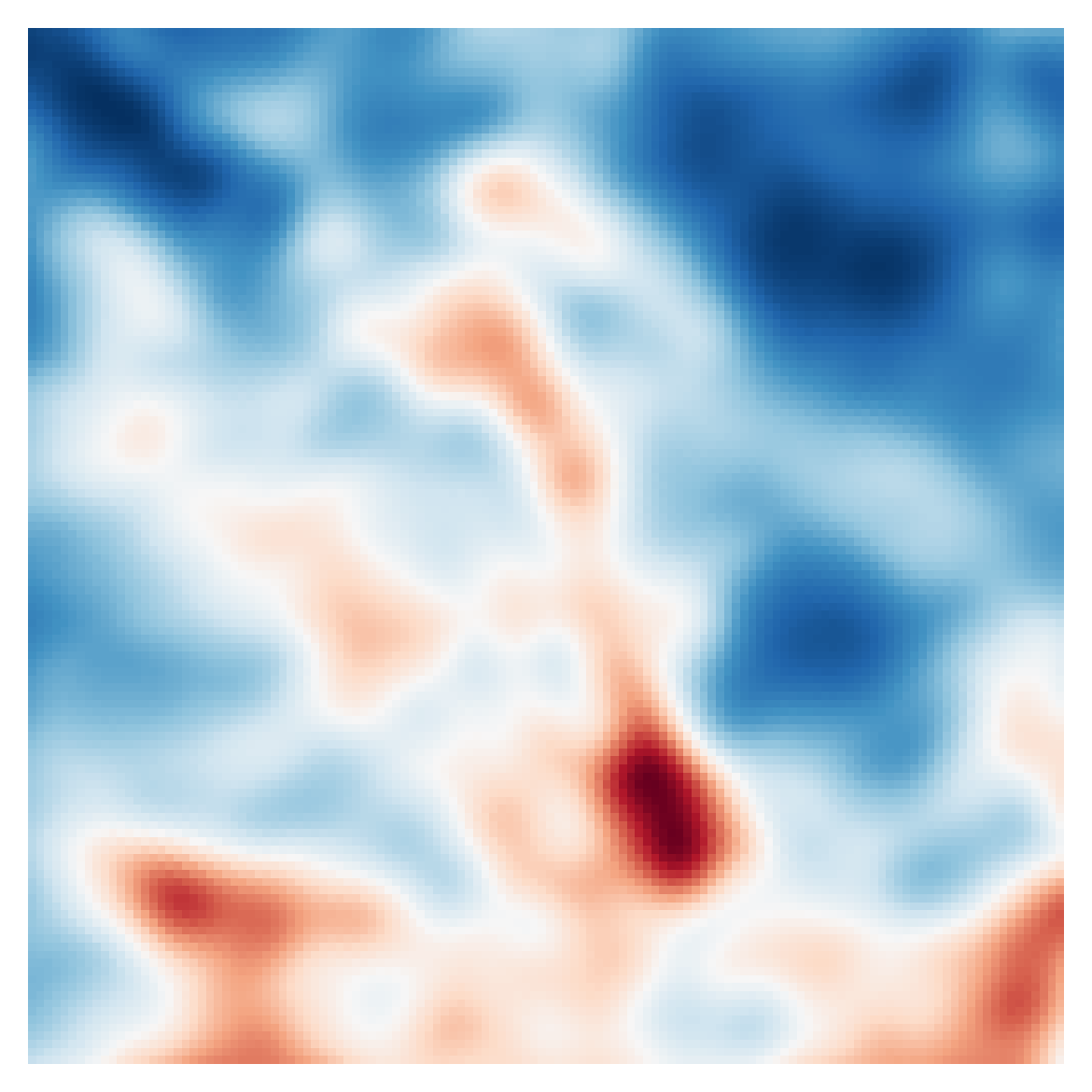} 
   &
   \includegraphics[height = 0.27 \linewidth]{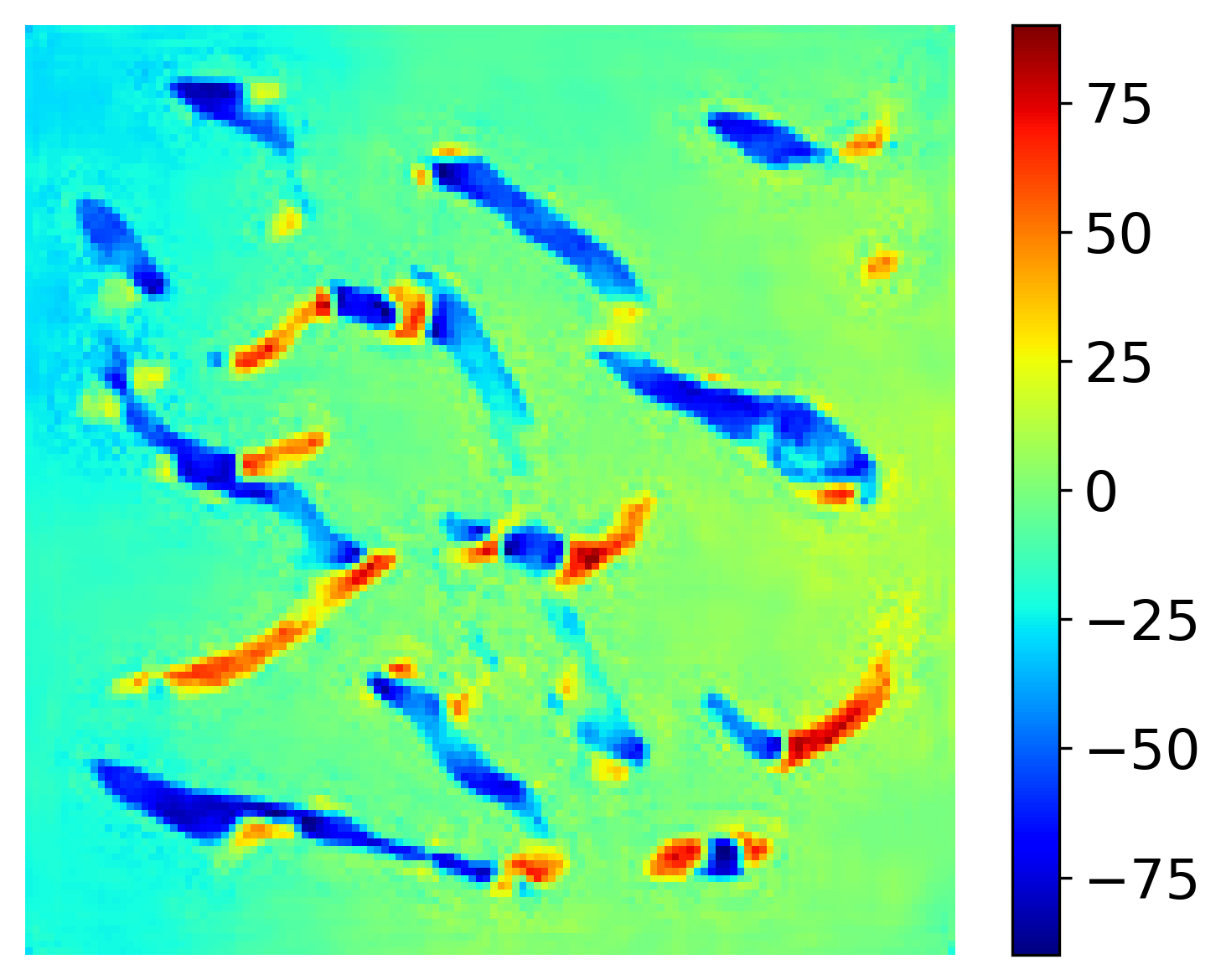}
   &
   \includegraphics[height = 0.27 \linewidth]{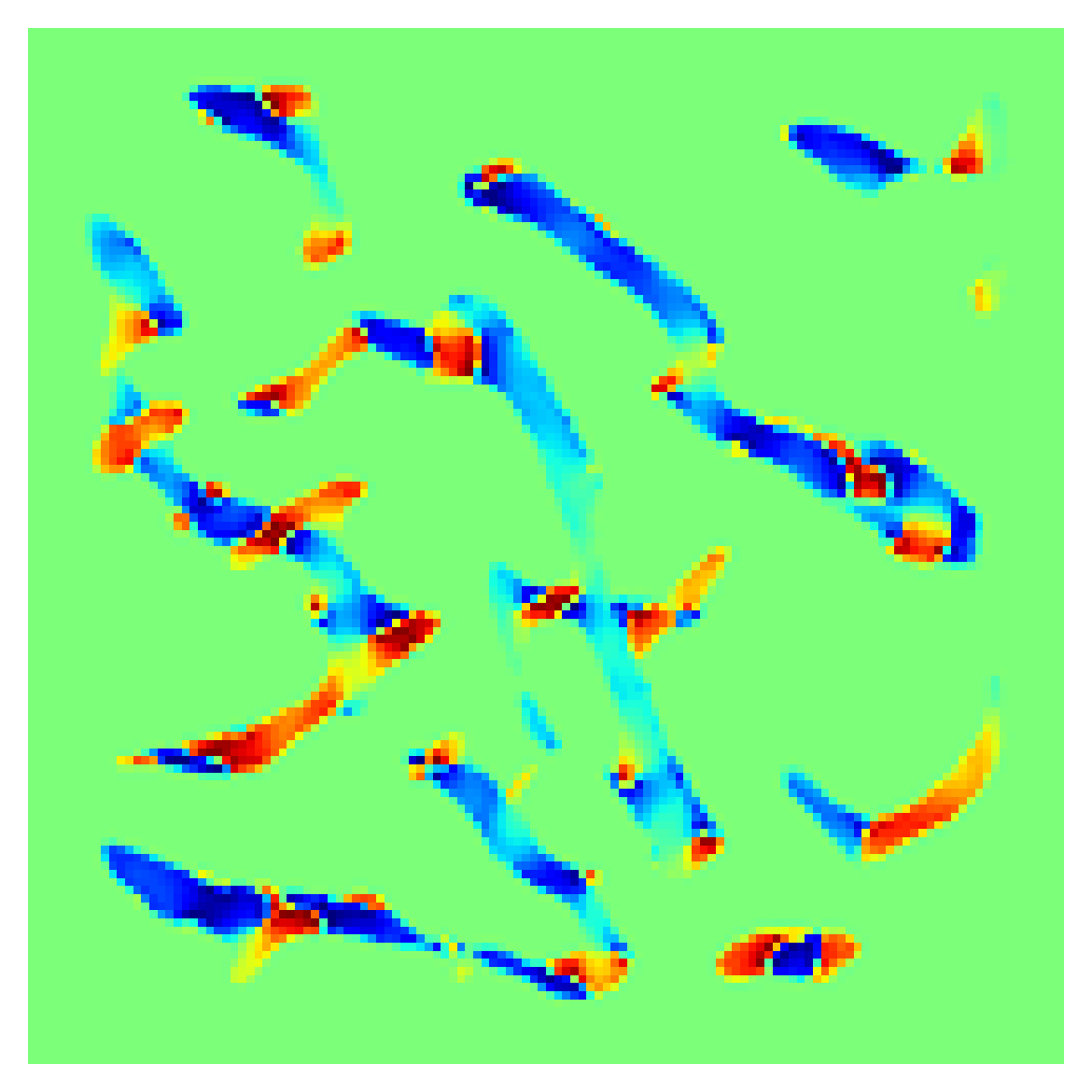} \\
    \small{} & \small{0.3001} & \small{0.2981} \\ %G82.65-2.00
   \includegraphics[height = 0.27 \linewidth]{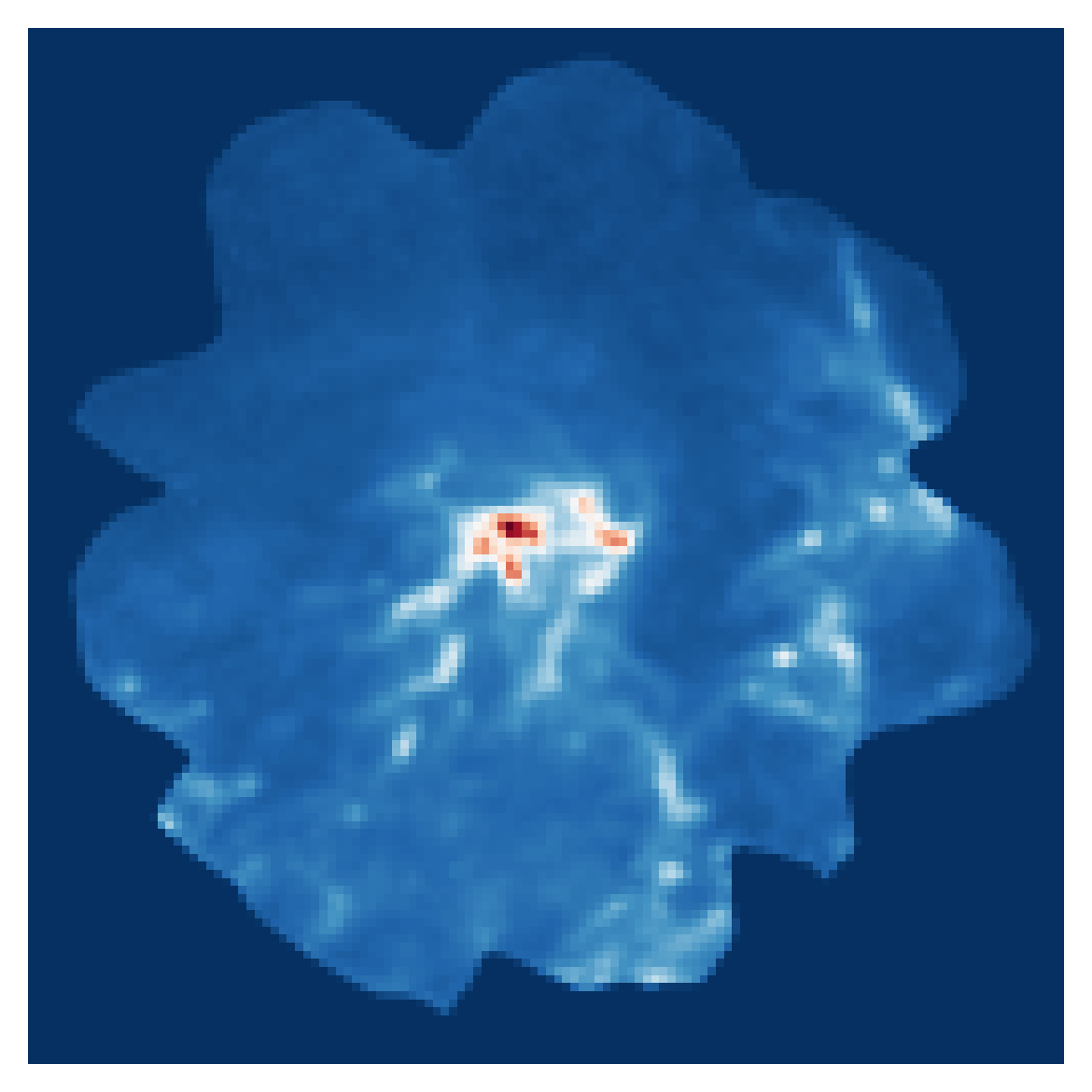} 
   &
   \includegraphics[height = 0.27 \linewidth]{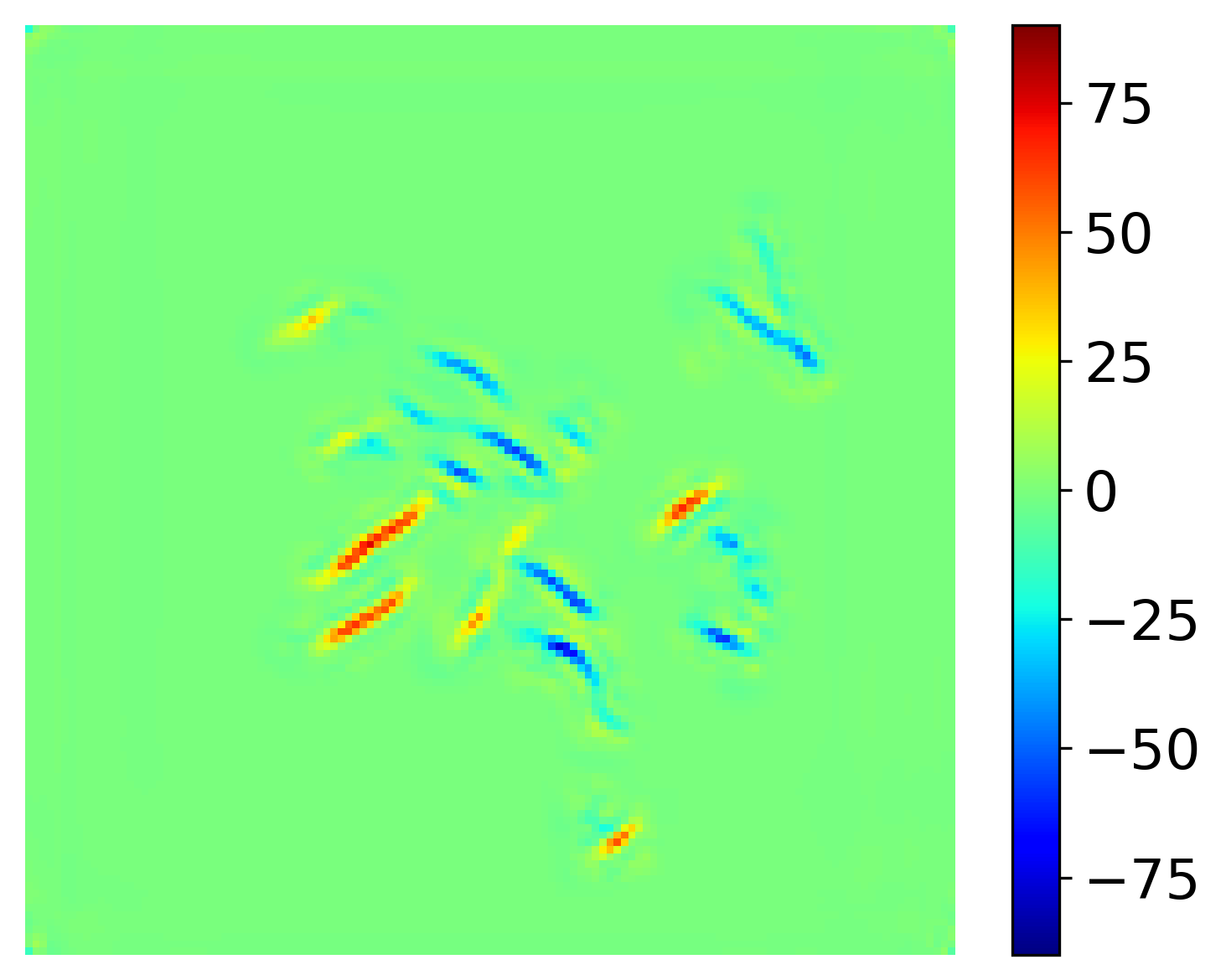}
   &
   \includegraphics[height = 0.27 \linewidth]{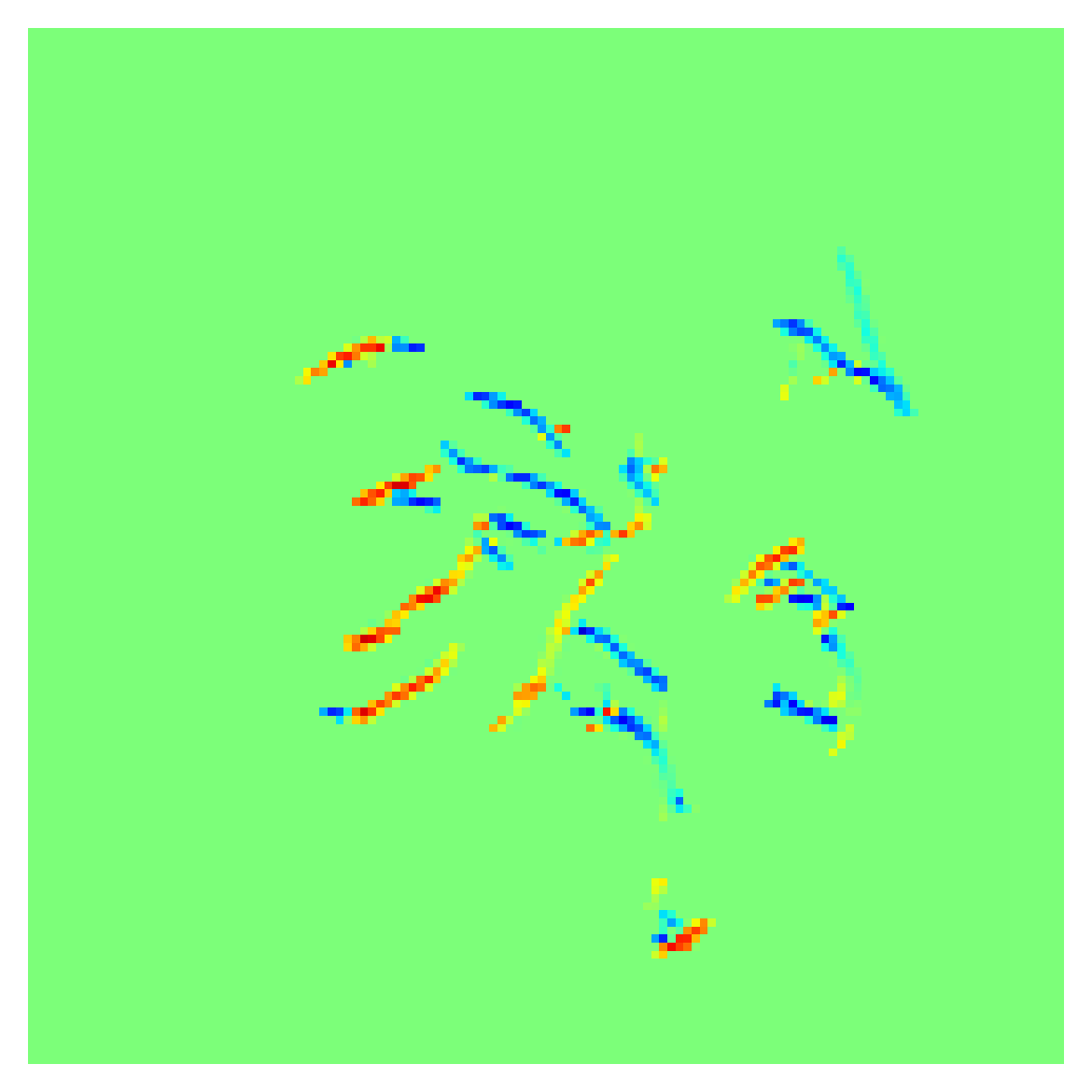} \\
      \small{} & \small{0.3001} & \small{0.2981} \\ %G82.65-2.00
   \includegraphics[height = 0.27 \linewidth]{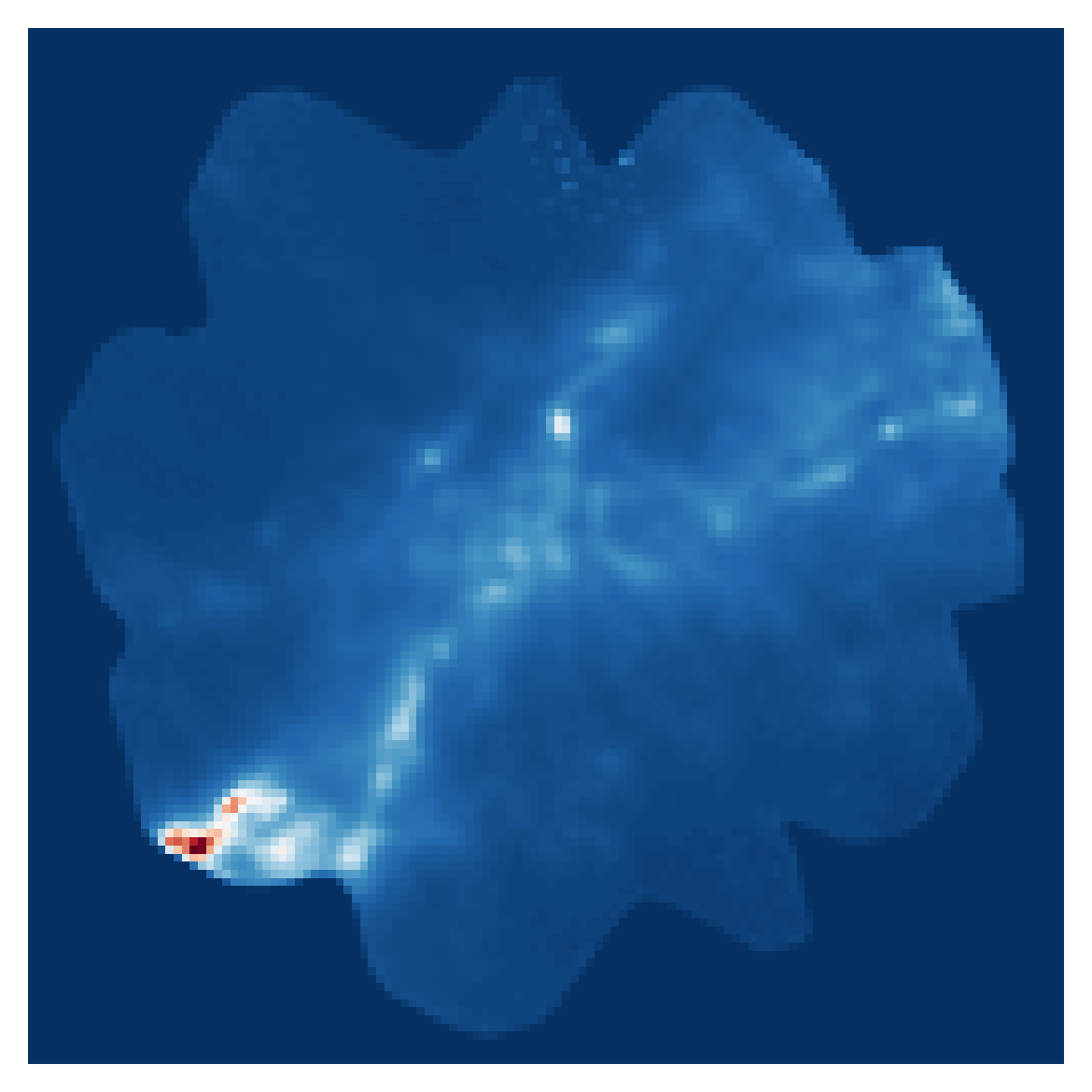} 
   &
   \includegraphics[height = 0.27 \linewidth]{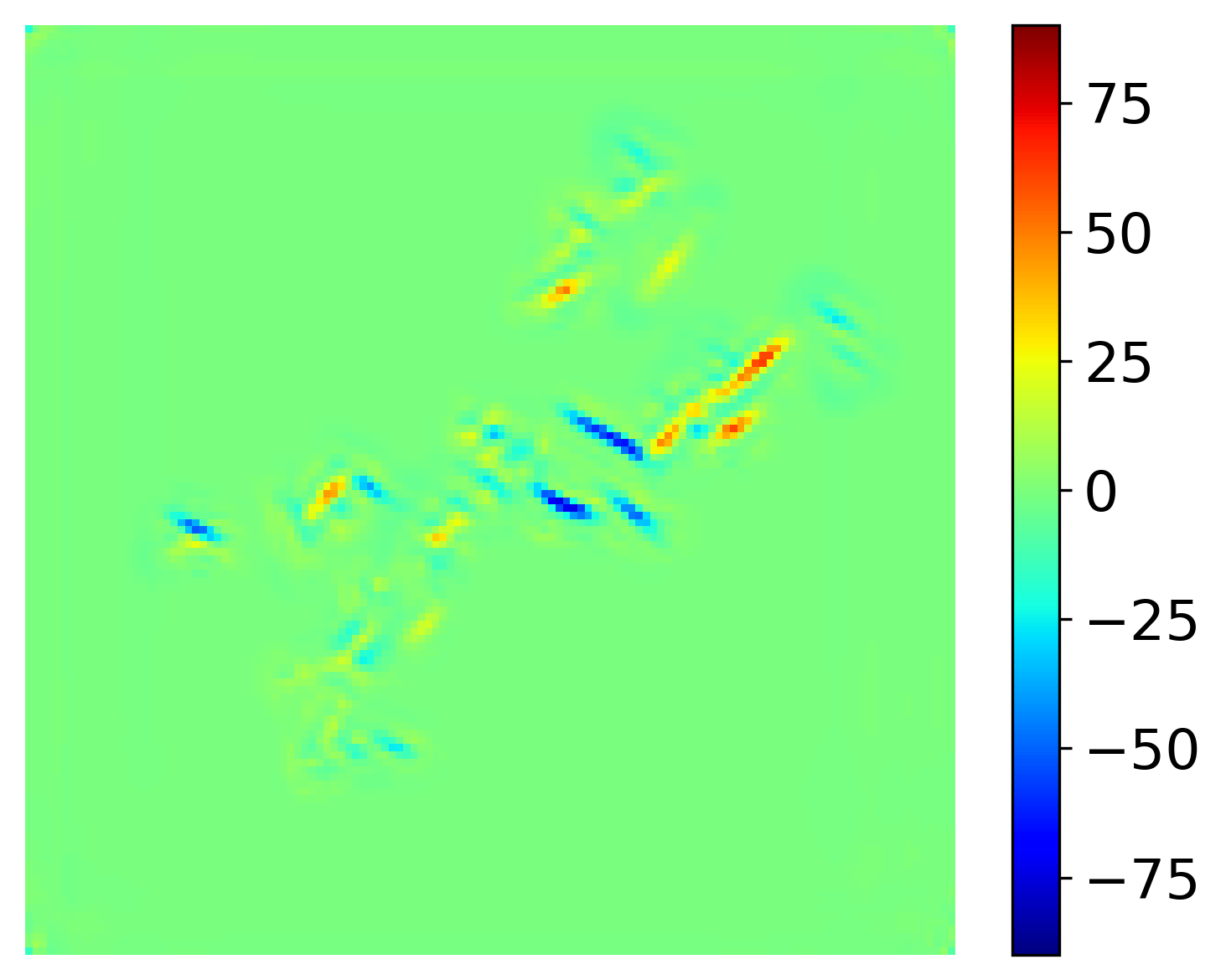}
   &
   \includegraphics[height = 0.27 \linewidth]{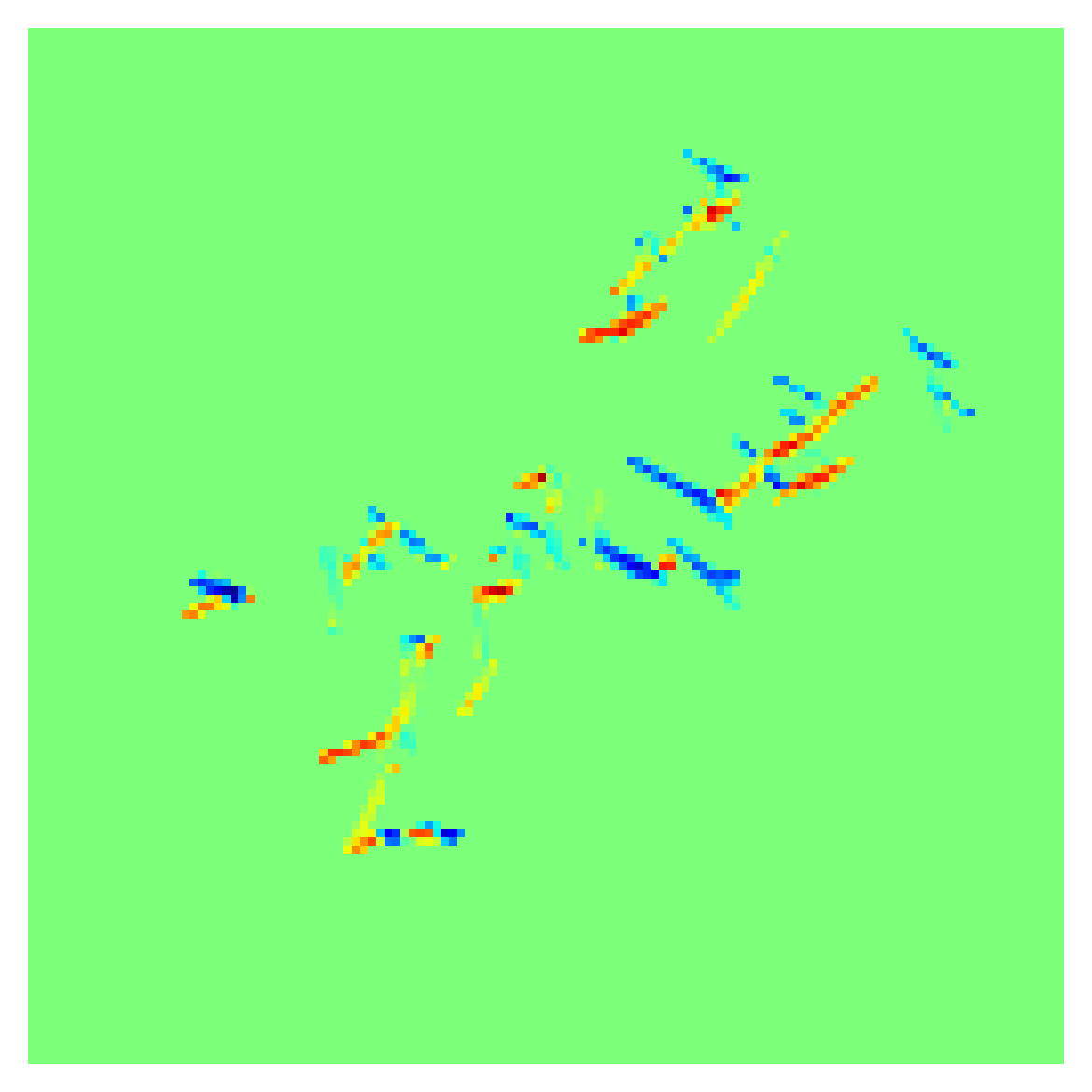} \\
       \small{} & \small{0.3001} & \small{0.2981} \\ %G82.65-2.00
   \includegraphics[height = 0.27 \linewidth]{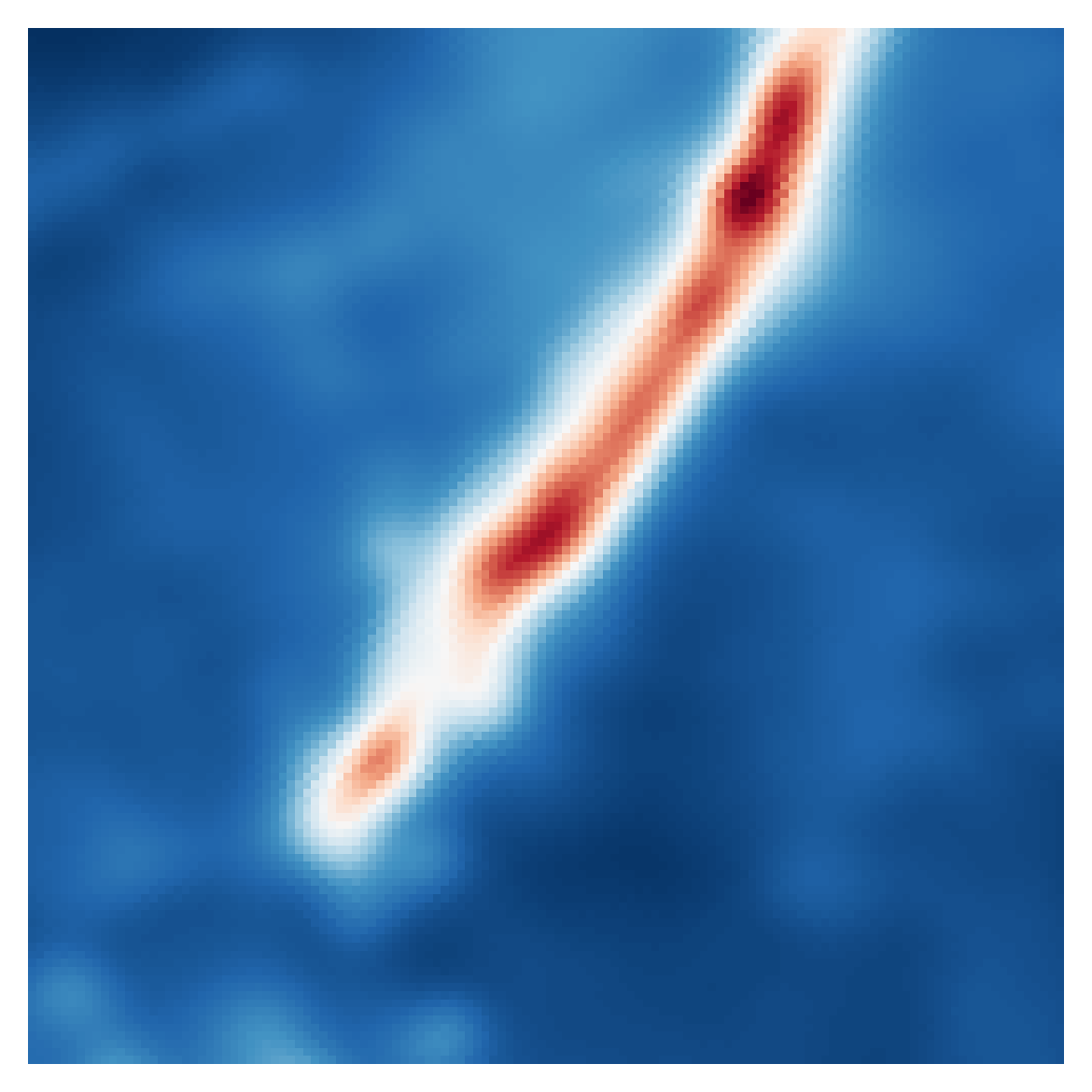} 
   &
   \includegraphics[height = 0.27 \linewidth]{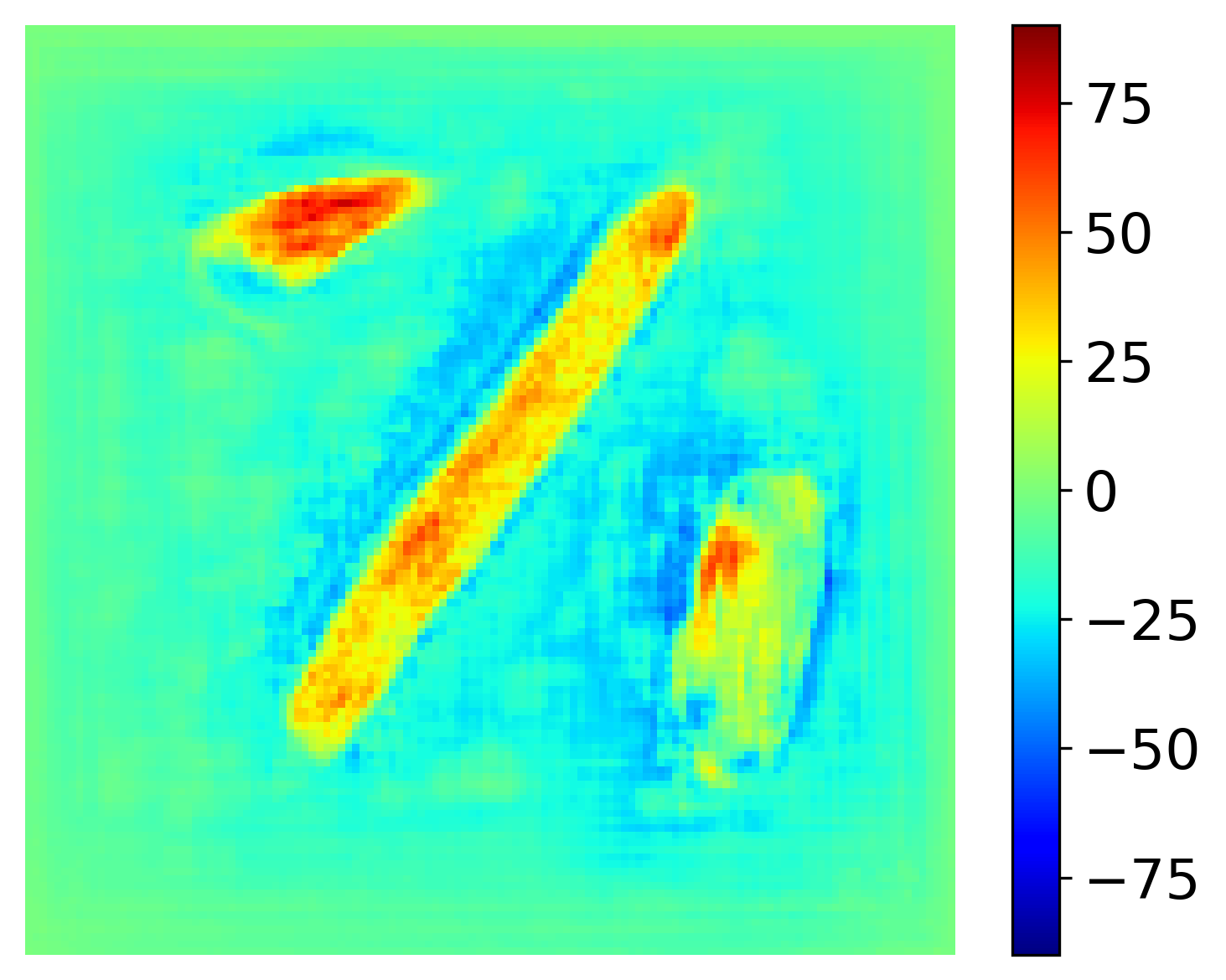}
   &
   \includegraphics[height = 0.27 \linewidth]{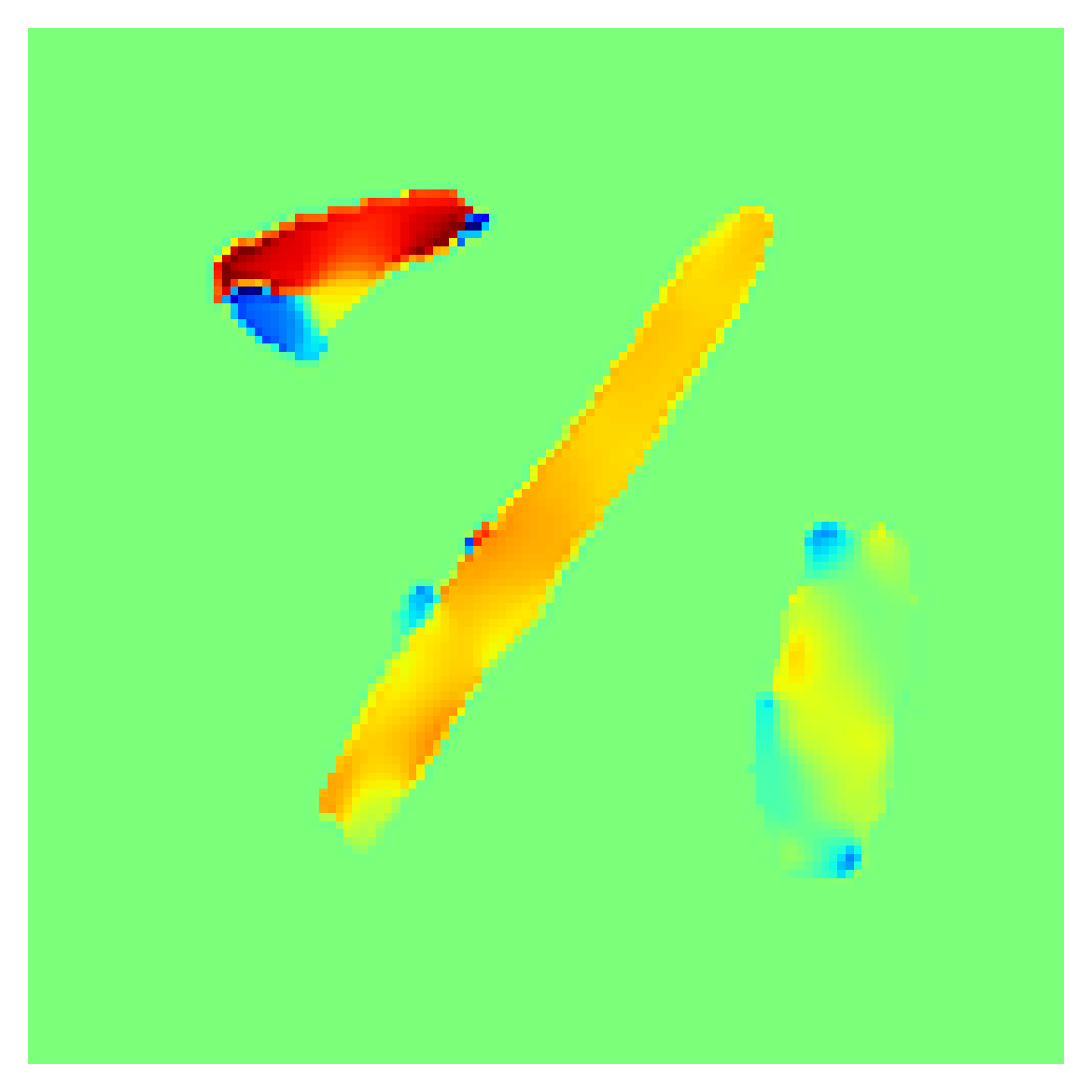}

\end{tabular}

    \caption{\quad Examples of filaments obtained using the U-Net model: the first column depicts the original intensity maps, the second column shows results of the U-Net, with the color code corresponding to the orientation angle in degrees, and the last column shows results of the RHT procedure. Numbers in the second and third columns indicate the structural similarity score measured by the MSSIM metric.}
    \label{fig:unet_examples}
\end{figure}

\subsubsection{Training}
In the training procedure, the input is the mask based on the RHT intensity result, similar to the masks used as the desired output for the Mask-RCNN. The output is the map with orientation angles, also provided by the result of the RHT. There, the orientation angles range between $-90^{\circ}$ and $90^{\circ}$ degrees with $0^{\circ}$ corresponding to the vertical. 
It is worth noting that the results of the Mask-RCNN model can be used as an input, however, this would require additional run of the RHT to determine the corresponding orientation angles. For simplicity and robustness, we opted to work with RHT results for the training.

Optimization of the parameters of the network is one of the essential steps in the training procedure. Due to a limited size of the training data sets, a particular attention was paid to the choice of the optimizer. We have tested different optimizers, among which the family of Adam optimizers showed the best training convergence and the loss results. Adam \cite{kingma2014adam} is an adaptive learning rate optimization algorithm that’s been designed specifically for training deep neural networks. 
%First published in 2014. The paper contained some very promising diagrams, showing huge performance gains in terms of speed of training.
Adam, Adamax, Adadelta and AdaGrad optimizers are among those that showed the best performance on our data sets. The respective initial values of the learning rate for these optimizers are $10^{-3}$, $5\times10^{-4}$, $5\times10^{-2}$, $10^{-3}$ with the corresponding loss values of 33.65, 119.65, 51.26 and 18.49. As a result of this search procedure, the AdaGrad optimizer achieved the best loss and the training performance, hence, it was chosen for the our task of finding the angle orientation of the filaments in the given dataset.

The U-Net neural network was trained on Planck-cc, \textit{Hershel}-based and HI4PI-based sample data sets. We excluded the Planck-1 sub-sample at this stage as it only contains a single filament and was already proven inefficient for Mask-RCNN. Additionally we resized the intensity maps and the output masks to $256\times256$ pixels. The training took 100 epochs. Similarly to Mask R-CNN, the Keras and TensorFlow deep learning packages with GPU support were used. The U-Net model in our implementation consists of ten modules, in which each module contains a convolutional, pooling/upsampling layers. The final layer has a linear activation function due to the regressive nature of the task.  
Minimum squared loss function was employed, with a learning rate of 0.0001. The minimum squared loss was measured at each training step to monitor the training process. The training has been shown to gradually converge after the 100 epochs of training.
The total number of trainable parameters of the network was 31,031,685.

\begin{table*}[]
\caption{\quad Estimation of performance of the U-Net training. Mean squared error measures the squared difference between the model's predictions and the desired output across the whole data set. The last column shows the mean difference between derived angles with the U-Net and with the RHT methods}
\centering
\setlength{\tabcolsep}{12pt}
\begin{tabular}{lllll}
\hline
\multicolumn{1}{|l|}{dataset} &  \multicolumn{1}{|p{2cm}|}{\centering train set sample \\ size (maps)} & \multicolumn{1}{|p{2cm}|}{\centering validation set sample \\ size (maps)} & \multicolumn{1}{|p{2cm}|}{\centering Mean squared error\\ (MSE) } & \multicolumn{1}{|p{2cm}|}{\centering Mean difference \\ (degrees)}\\ \hline
Hershel-based                            &  90                                                & 24                                                       & 267 & 2.68                                   \\
HI4PI-based                  &  120                                                 & 28                                                      &  187           & 1.35 \\
Planck-cc  &  100                                                & 37                                                      &  245 & 2.07
\end{tabular}
 
    \label{tab:unet}
\end{table*} 

\subsubsection{Results}
Results of the U-Net training for identification of orientation angles using three different data sets are presented in the Table \ref{tab:unet}. 
The mean squared error parameter (MSE) shows the squared difference between the U-Net model’s predictions and the ground truth, averaged across the whole data set.
The MSE will never be negative since we always squared the errors. The following equation formally defines the MSE:
\begin{equation}
MSE = \sum_{i=1}^{D}(Y_i-\hat{Y}_i)^2    
\end{equation}

U-Net allows to predict the orientation angles (see Fig.~\ref{fig:unet_examples}) with an accuracy comparable to RHT. This is confirmed by small average differences across each map. The mean over the average differences is shown in Table \ref{tab:unet}, which is of order of $1$ to $3$ degrees.

\section{Discussion}
\label{sec:discussion}

This section first describes the neural network models that we have additionally tested. Second, we summarize and discuss both our models.

\subsection{Alternative neural network models}
For orientation angle estimation, we have tried several approaches to test the most widely used models in machine learning. In particular, we compared the regular CNN model, the auto-encoder model, the decision tree regression, and U-Net models. 

The advantage of the CNN models is that they can capture spatial features in images, provide masks and perform reliable classification. However, they are not efficient in regression tasks. Thus, this type of neural network is not well suited for orientation angles determination. Hence, alongside CNNs, we used auto-encoder and U-Net models in our study. Due to the specific architecture, these models both use global and encapsulated local features in the images to provide a decoded 2D output with more precision. We also tried decision tree regression as one of the most robust and reliable classical machine learning methods. However, we faced over-fitting and poor generalization problem using this method. The sample variation significantly affected the results, in which some samples could reliably estimate the angles while other samples provided orders of magnitude less efficient orientation angle estimates.

\subsection{Mask R-CNN for filament identification} 

The Mask R-CNN-based model is used to distinguish filaments in the input maps. 
It is trained with data sets that contain structures of different sizes and morphological characteristics, with more "blobby", low angular resolution HI4PI data or more delicate, higher angular resolution \textit{Herschel} data (ratio of angular resolution of $20$). Using different data sets diversifies the learning procedure and ensures reliability. The output of the Mask R-CNN neural network consists of multiple layers, each containing a single filament. Furthermore, we can combine all masks to produce a single mask or perform a sorting procedure to limit the identified filaments in, e.g., a hierarchical order.  
Although the neural network was trained with maps containing ten filaments, the output contained more or less than ten significant filaments. This concludes that the neural network is able to make autonomous decisions. In addition, we showed that the training on a sample as small as 90 maps already gives results that are at least as reliable as the commonly used automated procedure such as the RHT. 
A comparison of the neural network results with the classic automated RHT procedure results in terms of morphological similarity showed that the neural networks approach provides outputs that are morphologically more representative of the original image.
In addition, once a neural network is trained, the computational time for a single map with around 500 pixels per side is less than 1 second compared to a few dozens of minutes with the RHT procedure on the same machine.

\subsection{U-Net for filament identification}

Although the U-Net architecture was previously used for the identification of thin filaments in microscopy, in this study, for the first time, it was used to determine the values of the orientation angle of the filaments. 

In principle, U-Net may be used to identify the orientation of filaments in a map containing masks of filaments obtained from any filament identification methods, such as intensity threshold or skeletons, among the most simple. Alternatively, it can be combined with more sophisticated procedures such as DisPerSE.

To find out how U-Net can identify orientation angles from the intensity maps directly, we tested various parameters of the network to estimate the best performance. This procedure was repeated on all data sets. We compared results of the U-Net with the following inputs: mask of the filament or the original intensity map. This allowed us to conclude that the latter gives significantly worse results. Hence, we propose using the architecture consisting of 2 stages: Mask-RCNN followed by U-Net to identify filaments and their orientation angles efficiently.

\section{Conclusion}
\label{sec:conclusion}
We applied machine learning approach to filament identification for the studies of the interstellar medium. The approach is based on neural networks and allows us to identify extended filaments of finite width and their orientation angles. To create training samples, we used a machine vision algorithm, the Rolling Hough Transform (RHT), that we applied to the publicly available astronomical data: the \textit{Planck} and \textit{Herschel} telescope and the HI4PI survey, which are the most used in ISM studies \citep{molinari2010,andre2010,clark2014,kalberla2016,rivera-ingraham2017,alina2019,carriere2019,planck2015-XXXII,planck2014-XXXV}. 
Our main goal was to find the best neural network architectures that would efficiently identify filaments and primarily estimate orientation angles. 

We found that two neural network models satisfy the required tasks: the Mask R-CNN and the U-Net. The first model is the best suited for filament mask construction while the second allows us to estimate the orientation angles of structures in the image. We recommend using a combination of the two models. 
However, it is worth noting that the U-Net model can be applied directly to the intensity or density image, and can also be used in combination with any mask of filaments. 

The main advantage of neural networks approach is that the models can be trained to identify structures of different sizes thus diminishing human bias. Such an approach minimizes parametrization, which facilitates application on large or diversified data sets. It opens an opportunity for neural network applications for relative orientation between interstellar filaments, hubs, and magnetic fields. Upon publication, the models set-up will be available via $\mathrm{GitHub}$ (https://github.com/danakz). In perspective, future work might be envisaged to improve the efficiency of the models regarding the set-up and the training samples.

\bibliographystyle{IEEEtran}
\bibliography{malefisenta}

%\clearpage
%\thispagestyle{empty}
\begin{IEEEbiography}[{\includegraphics[width=1in,clip,keepaspectratio]{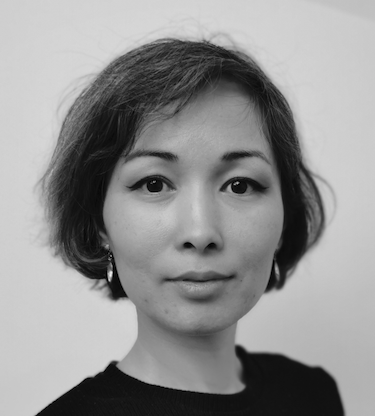}}]{Dana Alina} 
received the B.S. degree in Applied Physics from Pierre and Marie Curie University - Paris 6, Paris, France in 2009 and the M.S. degree in Astrophysics, Space Science and Planetology from Paul Sabatier University - Toulouse III, Toulouse, France, in 2011. She received the Ph.D. degree from Paul Sabatier University - Toulouse III University, Toulouse, France in 2015. The thesis covered analysis of dust polarised emission using the Planck satellite telescope data.

Since 2018 she has been working as the Provost Postdoctoral Scholar at the Physics Department at Nazarbayev University, Nur-Sultan, Kazakhstan. Her research interests include development of methods for data analysis in the field of polarimetric measurements, the dynamics of the interstellar medium and the magnetic fields.

Dr. Dana Alina was awarded the Bolashaq Fellowship by the Kazakhstan government in 2005.
\end{IEEEbiography}

%\parskip
\begin{IEEEbiography}[{\includegraphics[width=1in,clip,keepaspectratio]{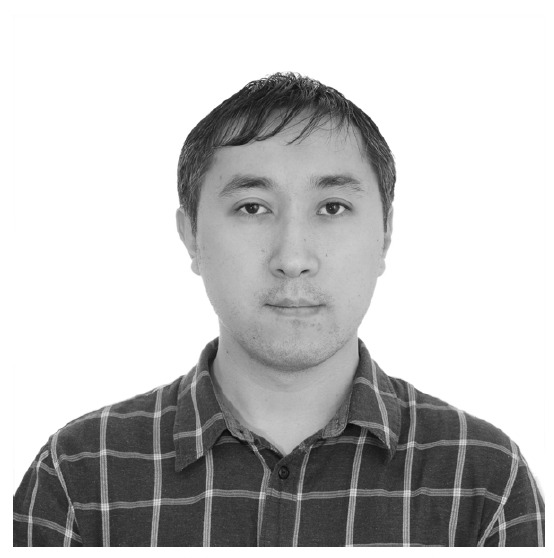}}]{Adai Shomanov} received the B.S. degree and M.S. degree in Computer Science from Al-Farabi Kazakh National University, Almaty, Kazakhstan, in 2010 and 2012, respectively. He received the Ph.D. degree in Computer Science from Al-Farabi Kazakh National University, Almaty, Kazakhstan, in 2018. The thesis covered the development of a scalable framework for parallel computations based on Mapreduce technology.

Since 2019 he has been working as a Research Assistant at the Department of Computer Science at Nazarbayev University, Nur-Sultan, Kazakhstan.
\end{IEEEbiography}

%\parskip
\begin{IEEEbiography}[{\includegraphics[width=1in,clip,keepaspectratio]{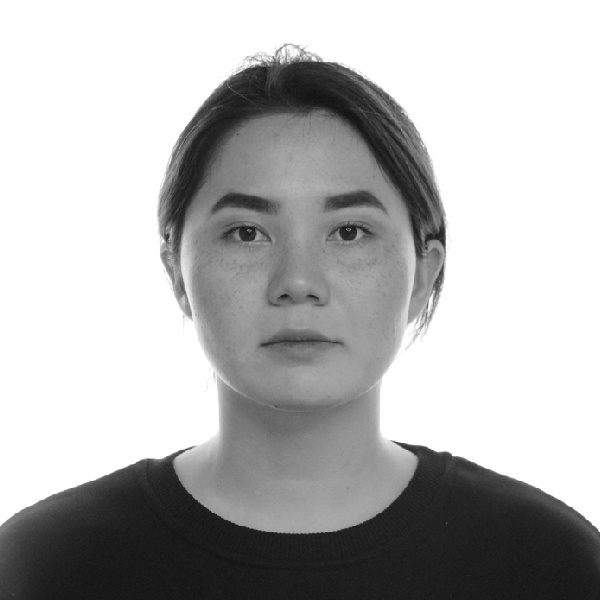}}]{Sarah Baimukhametova} received the B.S. degree in Mathematics from Nazarbayev University, Nur-Sultan, Kazakhstan, in 2021. She is currently earning her M.S. degree in Astrophysics and Cosmology from University of Padua, Padua, Italy.

From 2019 to 2021, she was a Research Assistant at Nazarbayev University, Nur-Sultan, Kazakhstan.

Sarah Baimukhametova was awarded the FLEX (Future Leaders Exchange) merit-based scholarship by the U.S. Department of State in 2016, the Nazarbayev University educational grant by the Kazakhstan government in 2017 and the academic scholarship by the Italian Ministry of Foreign Affairs and International Cooperation (MAECI) in 2021.
\end{IEEEbiography}

\EOD

\end{document}